\documentclass[11pt]{article}
\usepackage{amssymb}
\usepackage{xcolor}       
\usepackage{hyperref}
\hypersetup{
  colorlinks = true,      
  linkcolor  = blue,      
  citecolor  = blue,      
  urlcolor   = blue       
}
\usepackage[preprint]{acl}
\usepackage{xcolor}
\usepackage{graphicx}
\usepackage{times}
\usepackage{graphicx} 
\usepackage{subcaption} 
\usepackage{latexsym}
\usepackage{amsmath}
\usepackage{multirow}
\usepackage{svg}
\usepackage[T1]{fontenc}

\usepackage[utf8]{inputenc}

\usepackage{microtype}

\usepackage{inconsolata}
\usepackage{booktabs}
\title{Gradient Immunity: Null-Space Resistance to Malicious Fine-Tuning}

\author{
  Yuxuan Huang\textsuperscript{1,2},
  Xingyu Zeng\textsuperscript{3}$^{*}$,
  Tianhang Zheng\textsuperscript{4},
  Chaochao Lu\textsuperscript{1}\thanks{Corresponding authors.} \\
  \textsuperscript{1}Shanghai Artificial Intelligence Laboratory\\
  \textsuperscript{2}Shanghai Jiao Tong University\\
  \textsuperscript{3}Shenzhen University of Advanced Technology \\
  \textsuperscript{4}Zhejiang University \\
  \texttt{huangyuxuan@pjlab.org.cn, zengxingyu@suat-sz.edu.cn}\\
  \texttt{zthzheng@zju.edu.cn, luchaochao@pjlab.org.cn}
}

\begin{document}
\maketitle
\begin{abstract}
Released aligned large language models remain vulnerable to malicious downstream fine-tuning. Existing defenses are largely designed for the fine-tuning-as-a-service (FTaaS) paradigm or rely on downstream users to follow additional safety procedures, and therefore do not directly address the setting we study: a provider-controlled partially protected open-weight (PPOW) release setting in which most weights remain trainable while a small safety-critical component is preserved at release. We propose a \textbf{Unidirectional Safety Gate} (USG), instantiated as a \textbf{Null Space Cubic Layer} together with an \textbf{Inverse Adapter} inserted after the final Transformer layer. During downstream fine-tuning, the cubic layer suppresses or blocks gradients from harmful samples whose hidden states fall in a calibrated protected region, while the Inverse Adapter restores the base model's forward behavior. 
In practice, we calibrate a threshold using defender-held harmful data, allowing protection to generalize to nearby in-distribution harmful samples.
Across six evaluated model--dataset settings, USG keeps post-finetuning attack success rate close to
the pre-release level under a fixed release threshold, while maintaining high safe-pass rates on easier settings and exhibiting a clearer safety--utility trade-off on unsafe samples from BeaverTails. These results suggest that release-time representation-space blocking can raise the cost of malicious downstream adaptation without requiring downstream cooperation. The code is available at \url{https://github.com/OpenCausaLab/Gradient-Immunity}.

\end{abstract}

\section{Introduction}
Recent years have witnessed remarkable performance leaps in Large Language Models (LLMs), driven by architectural innovations like Mixture-of-Experts (MoE) \citep{cai2025survey,liu2024deepseek} and advanced reasoning techniques such as Chain-of-Thought (CoT) \citep{wei2022chain} and Reinforcement Learning with Verifiable Rewards (RLVR) \citep{su2025crossing,mroueh2025reinforcement}. These advances have significantly enhanced the reasoning capabilities of models, enabling them to achieve strong performance across downstream tasks such as mathematical reasoning, code generation, and causal understanding \citep{liu2024deepseek,grattafiori2024llama,achiam2023gpt,zhang2025causight,chen2026can}.

However, LLMs still face numerous safety challenges   \citep{lab2025safework,purpura2025building,qi2024safety,lin2025against,qi2026darwin}. Beyond the inherent
vulnerabilities during inference, such as insufficient depth in safety 
alignment \citep{qi2024safety,zhang2026internalizing,wu2026datashield} and susceptibility to misleading prompts that fail to counter 
toxic input \citep{melamed2024prompts,li2026evodefense,zeng2026trace,he2026early,he2026not}, they also exhibit significant weaknesses when confronted with
malicious AI trainers \citep{huang2024harmful}. Research has shown that a small amount of toxic training
data can completely break a model's original safety alignment \citep{wei2024assessing,wu2026datashield}.

Existing fine-tuning defense methods are primarily designed for the fine-tuning-as-a-service (FTaaS) paradigm \citep{huang2024harmful,weng2025think,li2026evodefense,zhang2026self,wen2026magic} and therefore do not directly match the provider-controlled PPOW release setting studied here, in which most weights remain trainable after release while a small safety-critical subset must remain protected under provider-side hardening \citep{qi2024evaluating}. We focus on low-budget downstream SFT under this same release assumption; FTaaS can be viewed as a special case in which stronger provider-side control makes the protected-subset assumption easier to satisfy. Current defense approaches can be categorized into three types \citep{huang2024harmful}: alignment-based methods \citep{rosati2024representation,huang2024vaccine,huang2024booster}, fine-tuning stage methods \citep{bianchi2023safety,zong2024safety}, and post-fine-tuning stage methods \citep{casper2024defending,yi2024safety}. Alignment stage defenses, such as Vaccine \citep{huang2024vaccine}, can only defend against specific types of data attacks and may weaken under stronger downstream adaptation. Fine-tuning stage defenses, like SafeInstr \citep{bianchi2023safety}, require the trainer to use a specific training framework; however, in this PPOW setting, a downstream fine-tuner may simply choose a different training pipeline for the exposed weights. Post-fine-tuning stage defenses, such as SafetyLock \citep{zhu2024locking}, require users to perform activation steering toward safe directions after training, which cannot be relied upon once the model is released. These methods therefore share a common limitation: they depend on downstream compliance with additional safety procedures \citep{huang2024harmful}. This motivates a built-in safety mechanism that remains active after release and reduces reliance on user cooperation.

To address this challenge, we propose a novel Unidirectional Safety Gate (USG), a mechanism intended for the PPOW setting rather than a fully productized hardening artifact. Inspired by null-space theory \citep{fang2024alphaedit}, our USG can be inserted as a plug-in module into a pretrained model. During downstream fine-tuning, it acts as a selective safety filter: for inputs that trigger harmful behaviors, their hidden states are projected toward a protected null-space region. Empirically, non-participating but in-distribution
harmful samples also tend to exhibit lower norm-ratio values than safe samples, which gives the defender room to choose a release-time threshold that extends blocking coverage beyond the null-space build set. This mechanism prevents gradient propagation from blocked harmful samples to earlier layers of the model, thereby reducing harmful pattern acquisition during downstream SFT.

The main contributions of this work are summarized as follows:
1) \textbf{Conceptual Framing}: We introduce the concept of Unidirectional Safety Gate (USG), a release-time blocking mechanism for the PPOW setting that reduces reliance on downstream user cooperation.
2)\textbf{Mechanism Design}: We instantiate USG with a Null Space Cubic Layer and an Inverse Adapter. The cubic layer induces a calibrated protected region for harmful representations and blocks harmful gradient propagation during downstream fine-tuning, while the Inverse Adapter restores the base model's forward behavior.
%
3) \textbf{Practical Modularity}: Our method introduces minimal parameter overhead and can be inserted as a plug-in module into compatible pretrained models. In our evaluated setting, it preserves safe utility while the norm-ratio statistics show that non-participating but in-distribution harmful samples also tend to fall closer to the protected region than safe samples, allowing threshold calibration to extend blocking coverage beyond the participating harmful samples. The present paper evaluates this blocking behavior directly rather than claiming a fully validated deployment artifact.

\section{Related Work}

\textbf{Fine-tuning Attacks and Defenses.} While the safety capabilities of LLMs during inference have become increasingly robust, their resistance to malicious fine-tuning remains insufficient \citep{qi2024evaluating}. In the last two years, several works have addressed this challenge under the fine-tuning-as-a-service (FTaaS) paradigm, e.g.,  \citep{rosati2024representation,huang2024vaccine,bianchi2023safety,zong2024safety,casper2024defending,yi2024safety,hu2025adaptive,jiang2025metadefense,liu2025dualguard,huang2024booster}. For example, \citet{huang2024booster} preserve alignment performance by attenuating harmful perturbations, whereas \citet{rosati2024representation} erase harmful representations to make them harder to recover during fine-tuning.

\textbf{Null Space Applications in LLMs.} Null-space methods have shown promising results in privacy protection \citep{cao2026osnip}, preserving unrelated knowledge during fine-tuning or knowledge injection \citep{fang2024alphaedit,hu2025alphafuse,sun2025mitigating}, mitigating catastrophic forgetting, and reducing hallucinations \citep{yang2025nullu}. To the best of our knowledge, our work is the first to apply null-space methods to fine-tuning defense.

\section{Problem Formulation}
\subsection{Threat Model}
\label{sec:threat_model}
We study a provider-controlled partially protected open-weight (\textbf{PPOW}) release setting. The provider releases a model in which most parameters remain open and trainable for downstream users, preserving standard fine-tuning capability, while a small safety-critical parameter subset
is preserved through provider-side hardening such as freezing, encryption, structural entanglement, or deployment-time fusion with downstream computation. The attacker has white-box access to the released architecture and full control over downstream training on the exposed weights, including the choice of data, optimizer, and training pipeline.

We exclude attacks that require deleting, bypassing, replacing, or independently retraining the protected component, because such attacks are substantially more expensive than ordinary downstream fine-tuning in the PPOW setting. Accordingly, we focus on jailbreak by parameter update on the exposed weights. Our experiments further restrict attention to downstream supervised fine-tuning (SFT) under a low-budget regime, with materially fewer resources than were used to train the released model, e.g. a single GPU with 80 GB memory and on the order of 100 harmful samples. We do not claim coverage of more general post-release optimization procedures. If an attacker had substantially greater resources, they could instead train or heavily repurpose a separate harmful model directly, making protection of the released model less relevant.

Let $M_0$ denote the released model with parameters $W_0 = (W_{\text{open}}^0, W_{\text{prot}})$, where $W_{\text{open}}^0$ are the exposed trainable weights and $W_{\text{prot}}$ is the protected safety-critical subset. We consider an adversarial trainer whose goal is to erode the safety alignment of $M_0$ through malicious downstream fine-tuning while preserving benign utility. We assume the adversary possesses a fully harmful QA dataset $\mathcal{D}_\text{adv}$ consisting of adversarial question--answer pairs designed to elicit prohibited behaviors (e.g. bypassing safety filters or leaking private data).

The attacker updates only the exposed weights, yielding a corrupted model $M_1$ with parameters $W_1 = (W_{\text{open}}^1, W_{\text{prot}})$:
\begin{equation}
    W_{\text{open}}^1 = \mathop{\arg\min}_{W_{\text{open}}} L\big(M(W_{\text{open}}, W_{\text{prot}}), \mathcal{D}_\text{adv}\big)
\end{equation}
where $L$ is typically the standard cross-entropy loss. The adversary’s objective is to maximize harmful response success on $\mathcal{D}_\text{adv}$ while retaining as much benign capability as possible. Formally, let $\text{IsToxic}(M,x) \in \{0,1\}$ be a binary indicator of whether model $M$ generates a harmful response to input $x$. A successful attack satisfies:

\begin{itemize}
    \item $\forall x \in \mathcal{D}_\text{adv}, \text{IsToxic}(M_0, x) = 0$ but $\text{IsToxic}(M_1, x) = 1$
    \item $M_1$ retains or improves performance on benign benchmarks compared to $M_0$.
\end{itemize}

\subsection{Defensive Objectives and Robust Alignment}
Our goal is to develop a proactive defense mechanism that transforms $M_0$ into a robust variant $M_0^{'}$ (with weights $W_0^{'}$) before it is released to downstream users. Let $D_{\text{toxic}}$ denote a poisoned dataset drawn from the same distribution as $\mathcal{D}_\text{adv}$. This defense aims to ensure that even if $M_0^{'}$ is subsequently fine-tuned on $D_{\text{toxic}}$ to produce $M_1^{'}$, the resulting model remains resilient to the adversarial influence of $D_{\text{toxic}}$ while maintaining performance on the safe dataset \(D_{\text{safe}}\). We assume that the model provider has a defensive dataset \(D_{\text{defense}}\), and \(D_{\text{toxic}}\cup D_{\text{safe}} \subseteq D_{\text{defense}}\).

Formally, we view the following as ideal design objectives for a release-time defense in the PPOW setting:
\begin{itemize}
    \item Safety Invariance (idealized): For adversarial inputs $x \in D_{\text{toxic}}$ with $\text{IsToxic}(M_0, x) = 0$, the desired outcome is that the safety state remains stable:
    \begin{equation}
        \text{IsToxic}(M_1^{'}, x) = \text{IsToxic}(M_0^{'}, x) = 0
    \end{equation}
    \item Utility Generalization (desired): The transformation from $M_0$ to $M_0^{'}$ should not hinder legitimate learning. For safe data, the model should still converge to an improved state:
    \begin{equation}
        \begin{aligned}
        \text{Perform}(M_1^{'}, \mathcal{D}_\text{safe}) &\geq \text{Perform}(M_0^{'}, \mathcal{D}_\text{safe})\\ &\approx\text{Perform}(M_0, \mathcal{D}_\text{safe})
        \end{aligned}
    \end{equation}
    \item Strict Constraint Enforcement (idealized): In the fully blocked regime, the defense would keep $M_1^{'}$ from becoming more harmful than $M_0$ on the protected harmful distribution, thereby reducing the adversarial utility of $D_{\text{toxic}}$ while preserving the released model's capacity for safe downstream adaptation.
\end{itemize}

\subsection{Formal Definition of the Unidirectional Safety Gate}
To achieve the transformation from \(M_0\) to \(M_0'\), we introduce a safeguard module into the original model.
Specifically, inputs from the safe dataset can pass through the safeguard structure normally, while the gradient backpropagation for inputs from the prohibited subset is blocked by this structure. The prohibited subset is denoted as \(\mathcal{D}_{\text{forbidden}}\).


\textbf{\textbf{Unidirectional Safety Gate}(USG) Definition}: Suppose a layer with parameters \(W\) is inserted after the k-th transformer layer. Let \(\mathcal{D}_\text{defense}\) be the input dataset and \(\mathcal{D}_\text{forbidden} \subseteq \mathcal{D}_\text{defense}\) be a subset of prohibited inputs. For any input sample $x \in \mathcal{D}_\text{defense}$, let $h_x=f^{(k)}(x)$ denote the hidden-state tensor at the input of this layer, where \(f^{(k)}(.)\) denotes the forward function of the embedding layer and the first \(k\) Transformer layers. Let the forward propagation output of this layer be denoted as \(f(h_x, W)\). During the training process of the network, let \(\mathcal{L}\) denote the overall loss function of the network. The gradient of the loss with respect to the layer input is \(\frac{\partial \mathcal{L}(h_x,W)}{\partial h_x}\).

\begin{figure}[t]
    \includegraphics[width=0.5\textwidth]{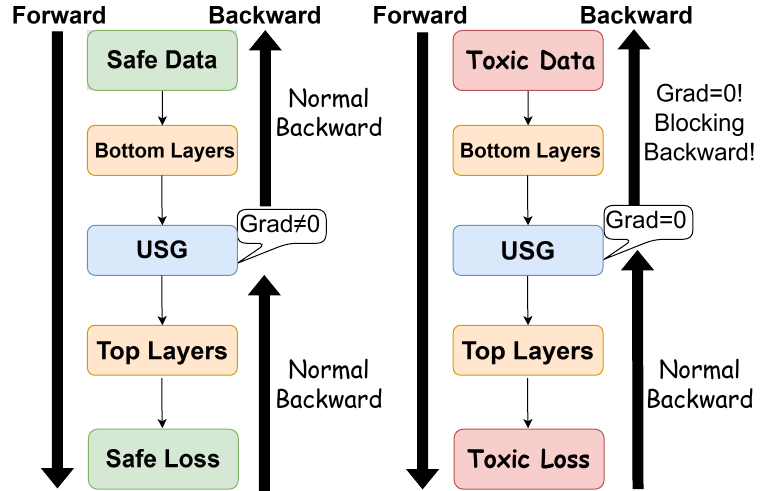}
    \caption{The figure illustrates the idealized behavior of a USG-style safeguard. For harmful data, it functions as a diode: during backpropagation, the USG sets gradients to zero, inducing gradient vanishing and thereby blocking gradient propagation. For safe data, it acts as a wire without inducing gradient vanishing in the idealized exact-blocking definition. Its core idea is the controllable induction of gradient vanishing—normally an undesirable behavior in model training.}
    \label{USG}
    \vspace{-10pt}
\end{figure}

If the following conditions hold:
\begin{itemize}
\item Forward Normality: For any input \(x \in \mathcal{D}_\text{defense}\), the forward computation \(f(h_x, W)\) is well-defined, producing a valid output tensor.

\item Backward Blocking: For any prohibited input \(x \in \mathcal{D}_{\text{forbidden}}\), we have
\begin{equation}
   \frac{\partial \mathcal{L}(h_x, W)}{\partial h_x} = 0.
\end{equation}
   For any input \(x \in \mathcal{D}_\text{defense} \setminus \mathcal{D}_{\text{forbidden}}\), no additional constraints are imposed on the gradient.
\end{itemize}
Then the layer is defined as a \textbf{USG} upon the prohibited input subset \(\mathcal{D}_\text{forbidden}\). It should be noted that USG is an abstract mechanism class rather than a specific architectural instantiation; the Null Space Cubic Layer and Inverse Adapter introduced later provide one concrete realization.
The role of USG in the model is as illustrated in Figure \ref{USG}.

\section{Methodology}

\subsection{Overall Framework}

To implement the aforementioned USG, we employ a cascade of a Null Space Cubic Layer and an Inverse Adapter. The cubic layer provides backward blocking on protected harmful inputs, while the Inverse Adapter compensates for the forward distortion and preserves the base-model behavior. In practice, exact null-space construction is approximated by a threshold on the norm ratio, implemented as a backward gate, rather than a hard semantic decision about whether a sample is harmful. The provider can calibrate a fixed harmful block-rate target on harmful reference data, and freeze a release threshold for deployment. Appendix~\ref{appendix:Threshold_Selection} gives the gate implementation details, and Appendix~\ref{appendix:safe_training_stability} reports the observation that, under a fixed threshold, safe fine-tuning still preserves a degree of harmful-blocking functionality.

\subsection{Cubic Layer}

As a concrete instantiation of the abstract USG, we introduce the \textbf{cubic layer}. 

The cubic layer is inserted after a Transformer layer, specifically after the residual connection of the preceding Transformer layer and before the input to the subsequent layer (which need not be a Transformer layer).

\begin{figure}[t]
    \includegraphics[width=0.5\textwidth]{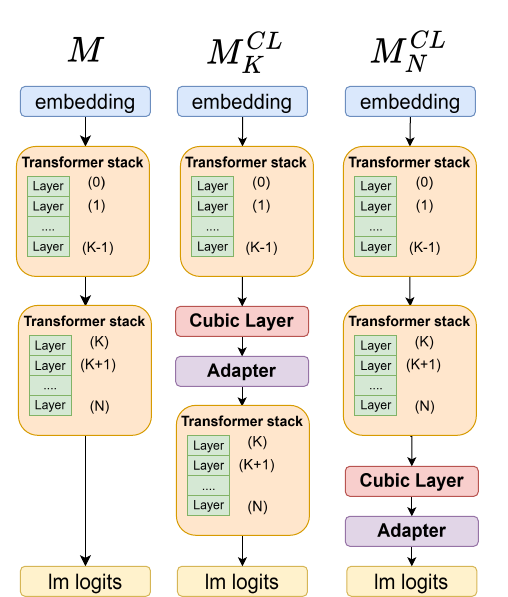}
    \caption{The roles of the Cubic Layer and Inverse Adapter in the model are illustrated in the figure. Here, \( M \) denotes the original model, \( M_{k}^{\text{CL}} \) represents the model obtained by inserting the Cubic Layer and the Inverse Adapter after the \( k \)-th Transformer, and \( M_{N}^{\text{CL}} \) represents the model obtained by inserting the Cubic Layer and the Inverse Adapter after the \( N \)-th Transformer.}
    \label{Cubic_layer}
    \vspace{-10pt}
\end{figure}

Suppose the cubic layer is inserted after the \(k\)-th Transformer layer; we denote it as \(CL_k\), where CL stands for Cubic Layer. Given a model \(M\) composed of \(n\) Transformer layers, if a cubic layer is inserted after its \(k\)-th Transformer layer, we refer to the resulting model as \(M^{\text{CL}}_{k}\). Let \(f^{(i)}\) denote the submodel of \(M\) consisting of the embedding layer and the first \(i\) Transformer layers, such that its output for an input \(x\) is \(f^{(i)}_{x}\). Clearly, because the first \(k\) Transformer layers of \(M^{\text{CL}}_{k}\) are identical to those of \(M\), the submodel formed by the first \(i\) Transformer layers (\(i \le k\)) of \(M^{\text{CL}}_{k}\) is also \(f^{(i)}_{x}\). Moreover, to avoid compromising the original performance of the model, an Inverse Adapter is appended after the cubic layer to analytically recover its forward output. The role of the Cubic Layer in the model is as illustrated in Figure \ref{Cubic_layer}.

\subsubsection{Definition}  

\textbf{Definition}. Let \( h \in \mathbb{R}^{B \times L \times D} \) be the layer-input hidden-state tensor, where
- \( B \) denotes the batch size (the number of independent samples processed in one forward pass),
- \( L \) denotes the sequence length (the number of tokens per sample, possibly after padding or truncation),
- \( D \) denotes the hidden dimension (the dimensionality of the feature vector for each token).

Let \( W \in \mathbb{R}^{D \times D} \) be a pre-set and unlearnable parameter. To facilitate the matrix operations defined below, we reshape this hidden-state tensor as \( h \in \mathbb{R}^{B \times L \times 1 \times D} \), i.e., we introduce a singleton dimension.

The output of the cubic layer is then defined as
\begin{equation}
O(h) = ((hW) h^\mathsf{T}) h,
\end{equation}  

In this paper, all matrix multiplications and the transpose operation $(\cdot)^\mathsf{T}$ are applied to the last two dimensions of the tensors, i.e., treating the first two dimensions $B$ and $L$ as batch dimensions via broadcasting. Specifically, for $h \in \mathbb{R}^{B \times L \times 1 \times D}$, its transpose $h^\mathsf{T} \in \mathbb{R}^{B \times L \times D \times 1}$ is obtained by swapping its last two dimensions. If \(W\) satisfies \(hW = 0\) and \(hW^\mathsf{T} = 0\), then this cubic layer is referred to as a \textbf{Null Space Cubic Layer}.


\subsubsection{Gradient of Cubic Layer}

Consider a backward pass with loss \(L\), and let \(\frac{\partial L}{\partial O(h)} = G_B\). Given the output of the cubic layer \(O(h) = ((hW) h^\mathsf{T}) h\), \(\frac{\partial L}{\partial h}\) can be decomposed into the following three terms.
\begin{equation}
\begin{aligned}
\text{term}_1 &= (G_B h^\mathsf{T}) (h W^\mathsf{T})\\
\text{term}_2 &= (h G_B^\mathsf{T})(h W)\\
\text{term}_3 &= ((h W^\mathsf{T}) h^\mathsf{T}) G_B
\end{aligned}
\end{equation} 

where \(\cdot\) denotes matrix multiplication. The total gradient is then given by

\begin{equation}
\frac{\partial L}{\partial h} = \text{term}_1 + \text{term}_2 + \text{term}_3.
\end{equation}

Detailed calculation is provided in Appendix \ref{sec:appendix}.

Observing \(\text{term}_1\), \(\text{term}_2\), and \(\text{term}_3\), we note that each contains the factors \(hW^\mathsf{T}\) and \(hW\). If \(W\) is such that both \(hW^\mathsf{T}\) and \(hW\) are approximately zero, the \textbf{vanishing gradient phenomenon can be induced}, thereby blocking backward propagation through the model. To achieve this effect, we construct a symmetric matrix \(W\) from the common right null space of the dataset \(D_{\text{forbidden}}\) for which gradient blocking is intended.

In other words, for layer-input hidden-state tensors \(h\) induced by samples in \(D_{\text{forbidden}}\), \(W\) is chosen so that \( hW = hW^\mathsf{T} = 0 \) in the exact setting. In our experiments, we use \( \|hW\| \) to quantify how close \( hW \) is to the zero vector, where \( \|\cdot\| \) denotes the Frobenius norm.

\subsubsection{Null Space Cubic Layer}

\textbf{Preliminary.} For a model \(M^{\text{CL}}_k\), let \(S_Y = \{ Y_1, Y_2, \dots, Y_n \}\) denote the tokenized inputs obtained from dataset \(D\), and define the set of layer-input hidden-state tensors

\begin{equation}
S_{f^{(k)}_Y} = \{ f^{(k)}(Y_1), f^{(k)}(Y_2), \dots, f^{(k)}(Y_n) \},
\end{equation}

By construction, \(S_{f^{(k)}_Y}\) is precisely the set of layer-input hidden-state tensors for the cubic layer in \(M^{\text{CL}}_k\).

Define \(h_j = f^{(k)}(Y_j)\) for \(j = 1, 2, \dots, n\), and let 
\begin{equation}
    S_h = \{h_1, h_2, \dots, h_n\}
\end{equation}

The right null space of a matrix \(H \in \mathbb{R}^{N \times N}\) is defined as

\begin{equation}
\operatorname{null}(H) = \{ p \in \mathbb{R}^N \mid H p = 0 \},
\end{equation}

and the common right null space of the set \(S_h\) of matrices is defined as
\begin{equation}
    \operatorname{null}_{\text{common}}(S_h)= \{ p \in \mathbb{R}^N \mid \forall h\in S_h, \, h p = 0 \}.
\end{equation}

\textbf{Definition.} A cubic layer is called a Null Space Cubic Layer if and only if

\begin{equation}
\forall h \in S_h, hW = hW^\mathsf{T} = 0.
\end{equation}

To construct such a cubic layer, we first compute a symmetric matrix \(W\) from the common right null space of \(S_h\). Specifically, let \(\operatorname{null}_{\text{common}}(S_h)\) denote the common right null space of \(S_h\).

Let \(V \in \mathbb{R}^{N \times r}\) be the matrix whose columns form a basis of \(\operatorname{null}_{\text{common}}(S_h)\), where \(r\) is the dimension of the common null space. Consequently, \(V\) satisfies

\begin{equation}
    \forall h \in S_h, \quad h V = 0.
\end{equation}

To ensure both \(hW = 0\) and \(hW^\mathsf{T} = 0\) for all \(h \in S_h\), we choose
\begin{equation}
    W = V V^\mathsf{T}.
\end{equation}
Since \(W\) is symmetric, we have \(W^\mathsf{T} = W\), and because the columns of \(V\) span the common right null space of \(S_h\), it follows that \(hW = h V V^\mathsf{T} = 0\) for every \(h \in S_h\).

Note that the parameter \(W\) of the cubic layer is \textbf{not learnable}; it is fixed and computed as described above.

\subsection{Inverse Adapter}
In the forward pass of the model \(M^{\text{CL}}_k\), the hidden states after the cubic layer differ from those in the original model \(M_0\); specifically, \(O(h) \neq h\) for the cubic layer. This mismatch renders the subsequent modules incompatible with the current output. Moreover, the parameter \(W\) of the cubic layer is fixed and not learnable. Therefore, we append an Inverse Adapter after the cubic layer to analytically undo this deterministic distortion and almost exactly recover the hidden states that would be produced without the cubic layer.

To derive this adapter, we first characterize the structure of the cubic-layer transformation. The output of the cubic layer has the form
\begin{equation}
O(h) = ((hW) h^\mathsf{T}) h.
\end{equation}

Since $h\in \mathbb{R}^{B\times L\times 1\times D}$ with matrix multiplications applied only on the last two dimensions, for each sample index $b\in{1,\dots,B}$ and token index $\ell\in{1,\dots,L}$ the layer acts independently on $h_{b,\ell}\in\mathbb{R}^{1\times D}$, yielding $O(h)_{b,\ell}=s_{b,\ell}h_{b,\ell}$ with $s_{b,\ell}=h_{b,\ell}Wh_{b,\ell}^{\mathsf{T}}$.

Hence, for every batch element and every token position, the cubic layer preserves the direction of the local hidden-state vector and only rescales it by a scalar determined by that local hidden state and $W$.

Since $W$ is positive semidefinite, this structure yields an analytic inverse in the non-degenerate case $s_{b,\ell} > 0$: if $y_{b,\ell}=O(h)_{b,\ell}$, then $y_{b,\ell}W y_{b,\ell}^{\mathsf{T}}=s_{b,\ell}^{3}$, and thus the inverse function is

\begin{equation}
h_{b,\ell}=y_{b,\ell} (({y_{b,\ell}W y_{b,\ell}^{\mathsf{T}}})^{-1/3}). 
\end{equation}

By contrast, when $s_{b,\ell}=0$, the inverse mapping at that position collapses to $y_{b,\ell}=0$ and becomes many-to-one, so there is no unique inverse.

The collapse cases are discussed in the Appendix~\ref{appendix:inverse_adapter}. Here, it suffices to note that the Inverse Adapter can be applied to any sample in our evaluated setting, compensating for the deterministic distortion introduced by the cubic layer without introducing additional trainable parameters.

\subsection{Combined USG Realization}
The proposed safeguard is formed by composing the Null Space Cubic Layer with the Inverse Adapter. Together, these two components
  realize the USG-style behavior studied in this paper: the cubic layer provides threshold-based backward blocking of gradient propagation
  for inputs from the prohibited subset \(\mathcal{D}_{\text{forbidden}}\), while the Inverse Adapter analytically compensates for the
  distortion introduced by the cubic layer.

\begin{itemize}
\item \textbf{Forward Normality:} For any layer-input hidden-state tensor \(h\), the forward computation remains well-defined. The cubic layer first applies the deterministic distortion \(O(h)=((hW)h^\mathsf{T})h\), and the Inverse Adapter then analytically compensates for this distortion. Thus, the combined cubic-layer-plus-adapter structure preserves the base-model forward behavior whenever the backward gate is inactive, up to negligible numerical error.

\item \textbf{Backward Blocking:} In the exact null-space formulation, if \(h=f^{(k)}(x)\) for some input sample \(x\in \mathcal{D}_{\text{forbidden}}\), then \(hW = hW^\mathsf{T} = 0\). Since \(\text{term}_1\), \(\text{term}_2\), and \(\text{term}_3\) all contain factors \(hW\) or \(hW^\mathsf{T}\), the backward gradient becomes exactly zero. In the practical instantiation used in this paper, we instead compute the norm ratio \(r(h)=\|hW\|/\|h\|\) and implement backward blocking through a threshold gate: gradients are suppressed when \(r(h)\le\tau\) and passed otherwise.
\end{itemize}

\subsection{Protected Release Assumptions and Tampering Resistance}

The threat model in Section~\ref{sec:threat_model} assumes that the protected component cannot be removed at negligible cost after release. Here we only note one plausible realization of that assumption: the provider may fuse the safeguard with adjacent computation so that isolating and deleting it requires additional reconstruction or retraining. We treat parameter fusion as one concrete supporting instantiation; Appendix~\ref{appendix:parameter_fusion} gives a theoretical discussion and a gating-based training method.

\subsection{Pipeline}

The proposed safeguard mechanism is implemented through the following steps:

\textbf{Step 1. Cubic Layer Insertion}. Starting from a pre-trained model \(M\), we insert a cubic layer followed by an Inverse Adapter after its last Transformer layer, resulting in the modified model \(M^{\text{CL}}_N\), where \(N\) denotes the number of Transformer layers in \(M\).

\textbf{Step 2. Collecting Hidden-State Tensors for the Cubic Layer}. Let \(\mathcal{D}_T\) be a set of harmful question--answer pairs collected by the model provider. The set of tokenized inputs obtained from \(D_T\) is \(S_Y = \{Y_1, Y_2, \dots, Y_n\}\). For every input \(Y \in S_Y\), we compute its layer-input hidden-state tensor for the cubic layer, i.e., its hidden representation after the \(N\)-th Transformer layer, by forward-propagating through the original model \(M\). This yields a set of hidden-state tensors
   \[
   \mathcal{S}_{f^{(N)}} = \{ f^{(N)}(Y_1), f^{(N)}(Y_2), \dots, f^{(N)}(Y_n) \},
   \]
   where \(f^{(N)}\) denotes the submodel consisting of the embedding layer and the first \(N\) Transformer layers of \(M\).

\textbf{Step 3. Null Space Extraction and Cubic-Layer Parameter Construction}. We compute the common right null space of the matrices in \(\mathcal{S}_{f^{(N)}}\), each treated as a row vector. Let \(V \in \mathbb{R}^{N \times r}\) be a matrix whose columns form a basis of this null space, where \(r\) is the dimension of the common null space. The parameter matrix of the cubic layer is then set to \(W = V V^{\mathsf{T}}\), which is symmetric and constructed from that common right null space, thereby ensuring that for any \(h \in \mathcal{S}_{f^{(N)}}\), we have \(h W \approx 0\) and \(h W^{\mathsf{T}} \approx 0\). We refer to harmful samples that participate in the null-space computation as \textbf{participating harmful samples}, namely harmful samples in \(D_{\text{forbidden}}\), and those that do not participate as \textbf{non‑participating harmful samples}, namely harmful samples in \(D_{\text{toxic}} \setminus D_{\text{forbidden}}\). For more details, refer to the corresponding section in the Appendix \ref{sec:appendix_impl}.

\textbf{Step 4. Obtaining the Inverse Adapter.} The Inverse Adapter is given directly by the analytic inverse function of the cubic layer, and therefore requires no iterative training. Its purpose is to preserve the base model's forward behavior on both safe and harmful samples, whereas the gradient blocking studied in this paper is entirely produced by the Null Space Cubic Layer.

\textbf{Step 5. Model Release}. After construction, the Null Space Cubic Layer and the Inverse Adapter are assumed to be preserved under the PPOW setting. One possible realization is discussed in
  Appendix~\ref{appendix:parameter_fusion}. The final model \(M_0'=M^{\text{CL}}_N\), equipped with the fixed cubic layer and the Inverse Adapter, is then released.

\section{Experiment}
We evaluate our method on two tasks: (1) resistance to harmful fine-tuning and (2) forward restoration by the Inverse Adapter. For the resistance task, we use a practical threshold-based instantiation of the null-space mechanism, implemented as a backward gate. To extend the sample limit, we use extended null-space selection; details are provided in Appendix~\ref{sec:appendix_impl}. Our main resistance results adopt a strict fixed-threshold 100\% harmful-block setting, where the threshold is calibrated once on defender-held harmful reference data and then kept fixed during evaluation.

Let $r(h)=\|hW\|/\|h\|$ and let $\tau$ denote the threshold. The \textbf{safe pass rate} is the proportion of safe samples satisfying $r(h)>\tau$, while the \textbf{harmful block rate} is the proportion of harmful samples satisfying $r(h)\le\tau$.

Further threshold-calibration details for the strict fixed-threshold 100\% harmful-block release setting are reported in Appendix~\ref{appendix:Threshold_Selection}.

\begin{table*}[t]
\centering
\small
\setlength{\tabcolsep}{5pt}
\begin{tabular}{llcccccc}
\toprule
\textbf{Model} & \textbf{Safeguard} & \multicolumn{2}{c}{\textbf{JailbreakBench}} & \multicolumn{2}{c}{\textbf{HarmfulBench}} & \multicolumn{2}{c}{\textbf{BeaverTails-H}} \\
\cmidrule(lr){3-4}\cmidrule(lr){5-6}\cmidrule(lr){7-8}
 &  & \textbf{Pre} & \textbf{Post} & \textbf{Pre} & \textbf{Post} & \textbf{Pre} & \textbf{Post} \\
\midrule
\multirow{5}{*}{\textbf{Qwen-14B}}
& None              & 0.04 & 0.55            & 0.05 & 0.29            & 0.07  & 0.215           \\
& Booster           & 0.04 & 0.16            & 0.05 & 0.09            & 0.07  & 0.30            \\
& RepNoise          & 0.04 & 0.53            & 0.05 & /               & 0.07  & 0.38            \\
& Antibody          & 0.04 & 0.04            & 0.05 & 0.19            & 0.07  & 0.135           \\
& Null Space Cubic  & 0.04 & \textbf{0.04}  & 0.05 & \textbf{0.05}  & 0.07  & \textbf{0.07}  \\
\midrule
\multirow{5}{*}{\textbf{Llama-8B}}
& None              & 0.01 & 0.74            & 0.04 & 0.37            & 0.035 & 0.78            \\
& Booster           & 0.01 & 0.62            & 0.04 & 0.10            & 0.035 & 0.435           \\
& RepNoise          & 0.01 & 0.08            & 0.04 & /               & 0.035 & /               \\
& Antibody          & 0.01 & \textbf{0.00}  & 0.04 & 0.13            & 0.035 & 0.07  \\
& Null Space Cubic  & 0.01 & 0.01            & 0.04 & \textbf{0.04}  & 0.035 & \textbf{0.035}            \\
\bottomrule
\end{tabular}
\caption{ASR before and after harmful fine-tuning under different safeguarding methods on JailbreakBench, HarmfulBench, and BeaverTails-H. \textit{Pre} refers to ASR before harmful fine-tuning, while \textit{Post} refers to ASR after harmful fine-tuning on the corresponding harmful dataset. Under the strict fixed-threshold 100\% harmful-block release policy used for Null Space Cubic, post-fine-tuning ASR remains at the pre-release level across all six settings, indicating that harmful downstream updates do not further erode the safeguard in this evaluated strict-threshold setting once the frozen release threshold fully blocks the participating harmful reference set. \textbf{Boldface} marks the best post-fine-tuning defense result (lowest ASR) within each model--dataset block. The symbol `/` denotes runs with severe generation-quality degradation, including substantial utility loss, nonsensical generations, or mode collapse, for which the resulting ASR is not reported as a meaningful comparison.}
\label{main_result}
\end{table*}

\begin{table}[t]
\centering
\small
\setlength{\tabcolsep}{4pt}
\resizebox{\columnwidth}{!}{%
\begin{tabular}{lccc}
\toprule
\textbf{Model} & \textbf{JailbreakBench} & \textbf{HarmfulBench} & \textbf{BeaverTails-H} \\
\midrule
\textbf{Qwen-14B}  & 100\% & 100\% & 79\% \\
\textbf{Llama-8B} & 100\% & 100\% & 69\% \\
\bottomrule
\end{tabular}%
}
\caption{Safe pass rate under the strict fixed-threshold 100\% harmful-block release policy used in Table \ref{main_result}. This policy preserves perfect safe pass rate in the JailbreakBench and HarmfulBench settings for both model families, but becomes visibly more conservative in the harder BeaverTails-H setting.}
\label{tab:strict_release_safe_pass}
\end{table}
\subsection{Setup}

\subsubsection{Models and Datasets}
We adopt Qwen3-14B (reported as \textbf{Qwen-14B} in tables) \citep{yang2025qwen3} and Llama-3.1-8B (reported as \textbf{Llama-8B} in tables) \citep{grattafiori2024llama} as our test models for both tasks.

\textbf{Restoration datasets.} We evaluate forward restoration on 2,000 samples from the \textbf{Alpaca} dataset \citep{taori2023stanford}.

\textbf{Resistance datasets.} We use harmful QA examples from \textbf{JailbreakBench} \citep{chao2024jailbreakbench}, \textbf{HarmfulBench} \citep{mazeika2024harmbench}, and \textbf{BeaverTails-H} \citep{ji2023beavertails}, where BeaverTails-H denotes
  the unsafe subset of BeaverTails in this paper. For JailbreakBench and HarmfulBench, we use 100 harmful training examples per dataset;
  these same 100 examples serve both as the participating harmful build set used to construct the Null Space Cubic Layer and as the
  downstream harmful fine-tuning set in the strict persistence experiments. For BeaverTails-H, we use 100 harmful training examples to construct the Null Space Cubic Layer, while the remaining 100 harmful samples do not participate in the null-space computation but are used for threshold calibration as non-participating harmful samples. This strict persistence setting matches the PPOW setting above, in which the attacker’s harmful training data are covered by the defender-held harmful reference set.
  
  For each harmful build set, the resulting Null Space Cubic Layer is inserted after the last Transformer
  layer to obtain a safeguarded model \(M^{\text{CL}}_{N}\). This yields three safeguarded models per base model family; with two base
  model families, Qwen and Llama, we obtain a total of six models.

\subsubsection{Baselines.}

\textbf{Restoration baselines.} The original pre-trained model serves as the baseline for the restoration task. Specifically, we compare the performance of our defended model against that of the original model, both evaluated on the same safe dataset without any fine-tuning.

\textbf{Resistance baselines.} Among different defense strategies, both fine-tuning-stage and post-fine-tuning-stage methods can be trivially circumvented by an attacker with full control over the training pipeline. Consequently, we select the model without any safeguard, as well as models protected by alignment-stage defense methods, including \textbf{Rep Noise} \citep{rosati2024representation}, \textbf{Booster} \citep{huang2024booster}, and the latest gradient-based alignment method \textbf{AntiBody} \citep{nguyen2026antibody}, as our baselines.

\subsubsection{Hyper-Parameters and Training Settings.}

For the resistance tasks, supervised fine-tuning (SFT) is conducted. The AdamW optimizer \citep{loshchilov2017decoupled} is employed with full-precision training. The learning rate is set to 4e-5, with a maximum gradient norm of 1.0 and a warmup ratio of 0.2.

\textbf{Restoration setting.} No restoration-stage parameter training is performed; the Inverse Adapter is applied directly through the analytic inverse map.

\textbf{Resistance setting.} We adopt the \textbf{Inverse Adapter} fixed throughout the resistance tests, while the Null Space Cubic Layer provides the blocking behavior. As a concrete low-budget attack instantiation, we fine-tune only the last five Transformer layers using a standard SFT objective. All other hyper-parameters remain identical to those used in the restoration setting.

\subsection{Resistance Results}

We report the attack success rate (ASR) of models trained on harmful datasets under different safeguard mechanisms. Concretely, we first apply a safeguard (e.g., Null Space Cubic or Booster) to a base model \(M_0\) (e.g., Qwen3-14B or Llama-3.1-8B) to obtain a safeguarded model \(M_0'\), and then measure its ASR before and after harmful fine-tuning. The column \textit{pre} denotes the ASR before harmful training, and \textit{post} denotes the ASR on the same dataset after training. For the strict fixed-threshold 100\% harmful-block release policy used by Null Space Cubic, this comparison should be read together with the safe pass rates: unchanged post-ASR with high safe pass indicates persistence of the safeguard rather than blanket refusal.

For the Null Space Cubic method, the main resistance experiments use the strict fixed-threshold 100\% harmful-block release setting. The threshold is chosen as the maximum observed ratio \(\|hW\|/\|h\|\) over defender-held harmful reference samples at release time, and is then kept fixed throughout downstream harmful fine-tuning and evaluation. Table \ref{main_result} reports the resulting pre/post ASR, and Table \ref{tab:strict_release_safe_pass} reports the associated safe pass rates under the same frozen threshold. Under this setting, Null Space Cubic keeps post-fine-tuning ASR at the pre-release level in all six model--dataset settings. The backward gate induced by this frozen threshold preserves 100\% safe pass rate for both Qwen-14B and Llama-8B on JailbreakBench and HarmfulBench, while BeaverTails-H is more conservative, reducing safe pass rate to 79\% for Qwen-14B and 69\% for Llama-8B.

\subsection{Restoration Results}

Since Alpaca is a question-answering dataset, we evaluate restoration quality using the QA-pair accuracy (ACC). The
results are shown in Table \ref{Restoration_Results}, where response correctness is judged by DeepSeek-V3.

Table \ref{Restoration_Results} shows that, in our evaluated setting, the Inverse Adapter preserves base-model performance on the restoration benchmark. This is consistent with the intended role of the Inverse Adapter: it restores the cubic-layer outputs back toward the base model's outputs, while the gradient blocking itself is carried out by the Null Space Cubic Layer.

To directly verify that the Inverse Adapter is not merely preserving one downstream benchmark but is almost exactly undoing the cubic-layer distortion on benign data, we further evaluate the Inverse Adapter in isolation. Concretely, we load the released safeguarded model, feed the original last-layer hidden states through the Inverse Adapter, and compare the recovered hidden states and logits against the base-model outputs on a 500-sample benign mixed evaluation set. The results in Table \ref{tab:inverse_alignment} show perfect top-1 token agreement for all six runs, hidden-state cosine similarity numerically indistinguishable from 1, and very small relative reconstruction error. Additional details on the Inverse Adapter implementation and its practical behavior are provided in Appendix~\ref{appendix:inverse_adapter}.

\begin{table*}[t]
\centering
\small
\setlength{\tabcolsep}{6pt}
\begin{tabular}{llc}
\toprule
\textbf{Model} & \textbf{Method} & \textbf{ACC} \\
\midrule
\textbf{Qwen-14B}  & Base Model      & 66.0\% \\
                    & Inverse Adapter & 66.0\% \\
\midrule
\textbf{Llama-8B} & Base Model      & 60.0\% \\
                    & Inverse Adapter & 60.0\% \\
\bottomrule
\end{tabular}
\caption{Restoration ACC on the Alpaca dataset. In our evaluated setting, the Inverse Adapter matches the original base-model performance on this benchmark, consistent with its role as an analytic forward-restoration map rather than a separate trainable component.}
\vspace{-10pt}
\label{Restoration_Results}
\end{table*}

\begin{table*}[t]
\centering
\small
\setlength{\tabcolsep}{6pt}
\begin{tabular}{llcccc}
\toprule
\textbf{Model} & \textbf{Safeguard Source} & \textbf{Top-1 Agr. $\uparrow$} & \textbf{Logits KL $\downarrow$} & \textbf{Hidden Cos. $\uparrow$} & \textbf{Hidden Rel. L2 $\downarrow$}\\
\midrule
\textbf{Qwen-14B}  & JailbreakBench & 1.000000 & 4.93e-06 & 1.000000000 & 5.91e-06 \\
                    & HarmfulBench   & 1.000000 & 2.10e-04 & 1.000000000 & 1.04e-03 \\
                    & BeaverTails    & 1.000000 & 1.91e-03 & 1.000000000 & 2.75e-03 \\
\midrule
\textbf{Llama-8B} & JailbreakBench & 1.000000 & 1.82e-10 & 1.000000000 & 1.50e-07 \\
                    & HarmfulBench   & 1.000000 & 2.88e-05 & 1.000000000 & 6.04e-05 \\
                    & BeaverTails    & 1.000000 & 1.70e-05 & 1.000000000 & 5.57e-05 \\
\bottomrule
\end{tabular}
\caption{Inverse-only fidelity on a 500-sample benign mixed evaluation set. For each released safeguarded model, we apply only the Inverse Adapter and compare the recovered outputs against the original base-model hidden states and logits. Across all six runs, top-1 token agreement is perfect, hidden-state cosine similarity is numerically indistinguishable from 1, and hidden relative L2 error remains very small, supporting the claim that the Inverse Adapter almost exactly undoes the cubic-layer distortion. Under the inverse-adapter-plus-extended-null-space regime studied here, strict zeros are practically absent, so this same forward-preservation property extends to essentially all samples while the gradient blocking itself remains the responsibility of the Null Space Cubic Layer.}
\label{tab:inverse_alignment}
\end{table*}



\subsection{Blocking Generalization and Conditional Persistence}
To better characterize the behavior of the safeguard on unseen harmful data, we distinguish two questions. The first is \textbf{blocking generalization}: whether a safeguard constructed from participating harmful samples, i.e., harmful samples in \(D_{\text{forbidden}}\), can still induce lower \( \|hW\|/\|h\| \) ratios on \textbf{non-participating harmful samples}, i.e., harmful samples in \(D_{\text{toxic}} \setminus D_{\text{forbidden}}\), drawn from the same harmful distribution. The second is \textbf{conditional persistence}: whether the safeguard remains effective under continued harmful retraining once some harmful samples escape blocking. These two questions are related but not identical.

\begin{figure}[t]
    \centering
    \includegraphics[width=1\linewidth]{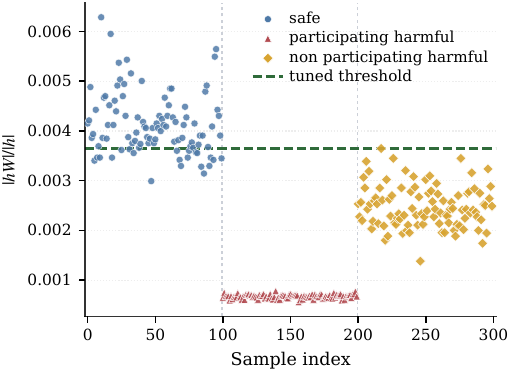}
    \caption{Norm-ratio visualization in the Qwen-14B BeaverTails-H setting of participating harmful samples \(D_{\text{forbidden}}\), non-participating harmful samples \(D_{\text{toxic}} \setminus D_{\text{forbidden}}\), and safe samples. Although the non-participating harmful samples are not used to construct the null space, they still tend to lie closer to the protected region than safe samples, showing a representation-level tendency that supports defender-calibrated threshold-based blocking beyond the build set.}
    \label{exp3_fig2}
    \vspace{-7pt}
\end{figure}

We first examine blocking generalization at a fixed release-time safeguard. Figure \ref{exp3_fig2} visualizes the norm ratio \(\frac{\|hW\|}{\|h\|} \) for different sample groups in the Qwen-14B BeaverTails-H setting. Although non-participating harmful samples are not used to construct the null space, their norm ratios still tend to be smaller than those of safe samples, indicating that the induced protected region extends beyond the participating harmful build set at the representation level. This is the main sense in which the null-space safeguard generalizes beyond \(D_{\text{forbidden}}\). It also clarifies why threshold calibration matters: because the defender controls the release-time threshold, this representation-level separation can be translated into practical blocking coverage on in-distribution harmful samples while preserving high pass rates on safe data.

We next consider a boundary case of conditional persistence under continued adaptation, in which the in-distribution harmful samples used for threshold calibration remain inside the protected region while a small fraction of out-of-distribution (OOD) harmful samples falls beyond the threshold. In this setting, the frozen threshold-controlled gate is typically preserved under small updates but can be breached by sufficiently large updates; details are deferred to Appendix~\ref{appendix:ood_coverage_stability}.

This yields a safety--utility trade-off through threshold calibration. Lower calibration thresholds reduce the risk of unnecessarily gating safe data but weaken harmful blocking coverage, whereas higher calibration thresholds improve harmful blocking at the cost of gating more safe data. USG should therefore be understood as a release-time calibrated threshold mechanism implemented as a backward gate, whose persistence depends on blocking coverage of harmful fine-tuning signals.

\section{Conclusion}

In conclusion, this paper proposes a Unidirectional Safety Gate implemented by a Null Space Cubic Layer together with an Inverse Adapter, which serves as a defense against malicious downstream fine-tuning in a PPOW setting by blocking the gradient backpropagation of harmful data inside a calibrated protected region under provider-side hardening assumptions. Experimental results across various datasets demonstrate two complementary findings: non-participating but in-distribution harmful data also tend to exhibit lower \(\|hW\|/\|h\|\) ratios than safe data, allowing appropriate defender-calibrated thresholds to extend blocking coverage beyond the participating harmful build set; and under the strict deployment-style setting in which the release threshold is calibrated on the complete defender-held harmful reference set available at release time, the safeguard can keep post-fine-tuning ASR at its pre-release level, although the accompanying safe-pass measurements show that this policy becomes more conservative in harder settings such as BeaverTails-H. In our evaluated restoration setting, the Inverse Adapter also preserves the base model's forward behavior and regains useful safe-side performance. We hope that this work will inspire future research on architectural safeguards for released LLMs that must remain trainable while preserving core safety guarantees.

\section*{Limitations}

\paragraph{Sensitivity of Null Space.}
The null space is sensitive to the input space; even minor perturbations can render it ineffective. Therefore, hard gradient blocking that relies on the null space, e.g., directly and completely blocking gradient backward, is highly susceptible to variations in the format and form of the input data. More precisely, the persistence of our safeguard depends on whether harmful training samples continue to be blocked. Like other representation-space defenses, USG relies on preserving a specific hidden-state geometry for harmful data. The main vulnerability arises when some harmful data fall outside the coverage of the defense corpus or otherwise escape blocking: those unblocked harmful updates gradually deform the original null-space structure used for blocking and can drive harmful representations back toward an attack-effective manifold. Soft gradient blocking based on the null space represents a promising direction for future research.

\paragraph{Dependence on Inverse-Branch Availability.}
Our strongest forward-preservation claim is tied to the inverse-adapter-plus-extended-null-space regime emphasized in this paper, where exact zeros are practically absent and the inverse branch therefore remains defined for essentially all samples. If a future deployment were to operate closer to the exact-null-space limit, then some hidden states could collapse to strict zero and the analytic inverse would no longer be uniquely defined at those positions. In that regime, preserving exact base-model outputs for every sample would require an additional design beyond the analytic inverse used here.

\paragraph{Limited Harmful Sample Number.}
As the number of harmful samples increases, the stacked harmful hidden-state matrix tends to approach full rank, so a non-trivial common null space may disappear. In that regime, the strict constraint used by our method can become infeasible. An Extended Null Space Selection strategy approximation instead of exact zero singular directions, see Appendix \ref{sec:appendix_impl} can partially mitigate this issue by replacing exact null-space constraints with approximate ones. Appendix~\ref{sec:appendix_impl} shows that exact null-space selection remains feasible for only tens of harmful samples in our evaluated settings, whereas extended null-space selection substantially enlarges the usable sample regime in the BeaverTails/Qwen case under a fixed 95\% harmful-block target. However, when the harmful sample size becomes sufficiently large, the resulting projected statistics of harmful and safe samples can overlap, making them difficult to separate reliably with a fixed threshold. Therefore, although Extended Null Space Selection improves robustness under moderate rank pressure, it does not remove the fundamental scalability limit: at large enough harmful-data scale, the method faces a decreasing margin between harmful and safe representations, which weakens discriminability and defense reliability.



\appendix

\section{Threshold Selection}\label{appendix:Threshold_Selection}

As discussed above, threshold selection mediates the trade-off between safety and continued trainability. Conceptually, the exact null-space analysis defines an ideal limit in which harmful samples satisfy \(hW=hW^{\mathsf{T}}=0\) and are therefore blocked by exact gradient cancellation. In the practical implementation, we realize this as a backward gate on the norm ratio \(r(h)=\|hW\|/\|h\|\):
\begin{equation}
    g(h) = \mathbf{1}[r(h) \le \tau], \qquad \frac{\partial \mathcal{L}}{\partial h} \leftarrow (1-g(h))\frac{\partial \mathcal{L}}{\partial h}.
\end{equation}
This threshold mechanism is an operational approximation to the ideal exact-null-space case for two concrete reasons: first, exact common-null-space construction becomes capacity-limited as the harmful sample set grows as mentioned above; second, finite-precision computation leaves numerically small but non-zero residual gradients, and these can still be amplified during backpropagation. The optimal threshold therefore depends on the model architecture, the harmful-data distribution, and the difficulty of the safe/alignment data. For example, using Alpaca as the safe dataset yields different behavior from using safe samples from BeaverTails, which is more challenging and therefore shifts the operating threshold between safe and harmful samples.

Our intended deployment protocol is not to keep retuning the threshold after release. Instead, the provider uses pre-release safe/alignment fine-tuning as a calibration stage and fixes a stable release threshold before deployment; the threshold is implemented post-release as a backward gate rather than an online decision rule.

\begin{figure*}[htbp]
    \centering
    \begin{subfigure}[b]{0.45\textwidth}
        \includegraphics[width=\textwidth]{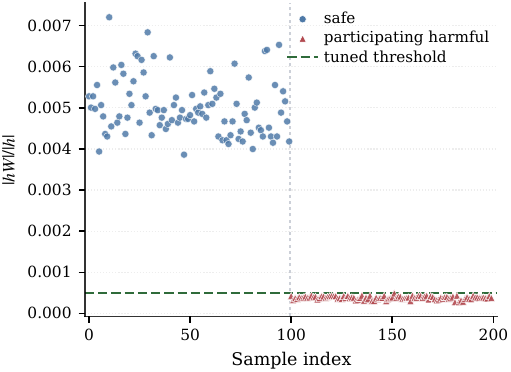}
        \caption{Qwen3-14B Threshold on JailbreakBench}
        \label{fig:sub1}
    \end{subfigure}
    \hfill 
    \begin{subfigure}[b]{0.45\textwidth}
        \includegraphics[width=\textwidth]{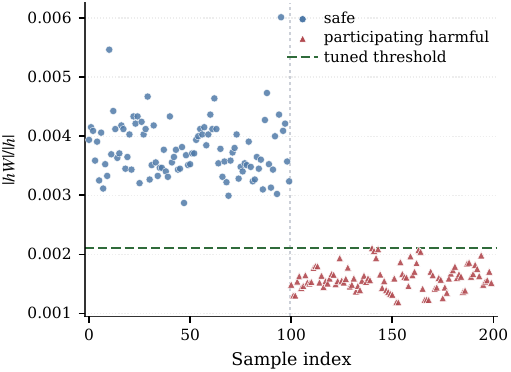}
        \caption{Qwen3-14B Threshold on HarmfulBench}
        \label{fig:sub2}
    \end{subfigure}
    \vskip\baselineskip 
    \begin{subfigure}[b]{0.45\textwidth}
        \includegraphics[width=\textwidth]{harmful_fig/qwen/b_step_0_ratio_scatter.pdf}
        \caption{Qwen3-14B Threshold on BeaverTails-H}
        \label{fig:sub5}
    \end{subfigure}
    \hfill
    \begin{subfigure}[b]{0.45\textwidth}
        \includegraphics[width=\textwidth]{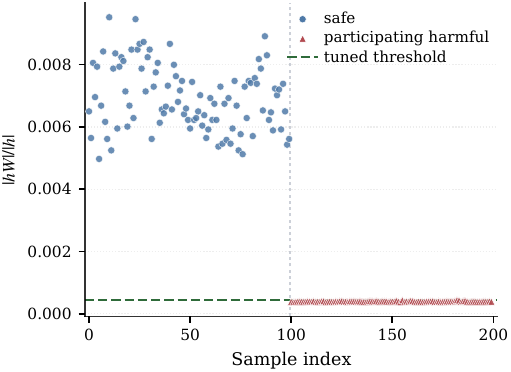}
        \caption{Llama-3.1-8B Threshold on JailbreakBench}
        \label{fig:sub3}
    \end{subfigure}
    \vskip\baselineskip 
    
    \begin{subfigure}[b]{0.45\textwidth}
        \includegraphics[width=\textwidth]{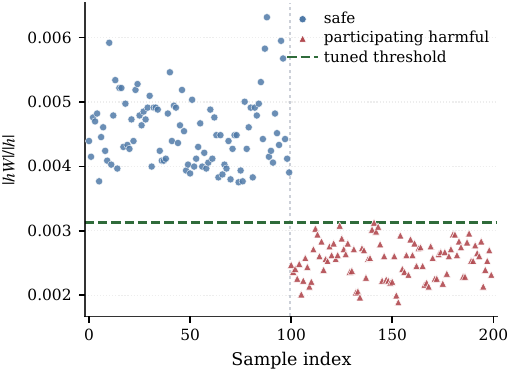}
        \caption{Llama-3.1-8B Threshold on HarmfulBench}
        \label{fig:sub4}
    \end{subfigure}
    \hfill
    \begin{subfigure}[b]{0.45\textwidth}
        \includegraphics[width=\textwidth]{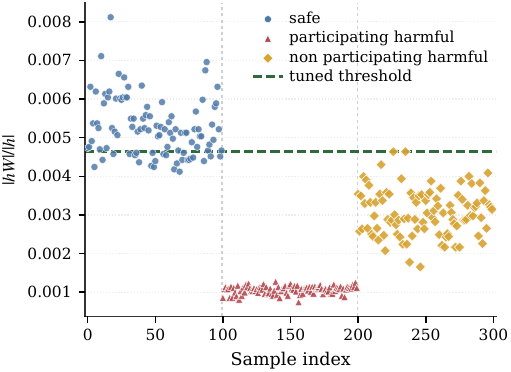}
        \caption{Llama-3.1-8B Threshold on BeaverTails-H}
        \label{fig:sub6}
    \end{subfigure}
    \caption{Visual Threshold Selection}
    \label{fig:visual}
\end{figure*}

In our experiments, we assess threshold quality using safe pass rate. During calibration, thresholds are selected by a harmful-ratio rule on the defender-held harmful reference set. Let \(r_1\leq r_2\leq \cdots \leq r_n\) denote the sorted norm ratios \(\|hW\|/\|h\|\) of the harmful reference samples at a given checkpoint. For a target harmful block rate \(b\in(0,1]\), we choose the calibration threshold as the empirical harmful quantile \(\tau_b = r_{\lceil bn\rceil}\), so that approximately a fraction \(b\) of harmful reference samples activates the backward gate. The strict 100\% release policy used in the main text is the limiting case \(b=1\), i.e., the threshold is set to the maximum observed ratio over all defender-held harmful reference samples, including harmful samples that did not participate in constructing the null space. The resulting 100\%-block calibration thresholds are shown in Table \ref{threshold_selection}, and the corresponding visualizations are shown in Figure \ref{fig:visual}. These thresholds should be interpreted as intermediate operating points produced by the pre-release calibration procedure rather than as final release thresholds.

The visualizations further show that BeaverTails is the most challenging setting: as the harmful sample set grows, harmful and safe statistics become harder to separate, and more held-out harmful samples approach the safe-side threshold region. This marks an upper-bound regime for blocking generalization under a fixed threshold and motivates the extended-null-space analysis discussed above.

\section{Null Space Stability during Safe Training}\label{appendix:safe_training_stability}

As discussed earlier, the persistence of our method under harmful retraining depends on whether harmful samples continue to be blocked: if all harmful data are blocked, the safeguard can remain effective, whereas failure arises when some harmful samples escape blocking and inject gradients. We therefore do not revisit harmful fine-tuning here. Instead, we study the complementary question that is important for practical downstream use: whether continued \emph{safe} training, unrelated to harmful content, also breaks the alignment induced by the cubic layer.

To answer this question, we continue safe training on Alpaca in a front-Transformer setting in which the Inverse Adapter remains frozen, only the last five Transformer layers are updated, and the cubic layer is inserted after the last Transformer layer. Training runs for 4 epochs with a learning rate of \(4\times10^{-5}\), cosine scheduling, 3\% warmup, no weight decay, gradient clipping at 1.0, BF16 precision, a per-device batch size of 1, gradient accumulation of 8 steps, and a maximum sequence length of 1024. We monitor the model every 20 training steps. We fix the threshold on the norm ratio \(\|hW\|/\|h\|\) to obtain a 95\% harmful block rate at the beginning and then evaluate the resulting safe pass rate in Alpaca. We use JailbreakBench, HarmfulBench, and 1000 samples from BeaverTails-H as harmful reference datasets to assess whether harmful and safe samples remain distinguishable throughout continued benign training.

\begin{figure*}[t]
    \centering
    \begin{subfigure}[t]{0.32\textwidth}
        \centering
        \includegraphics[width=\linewidth]{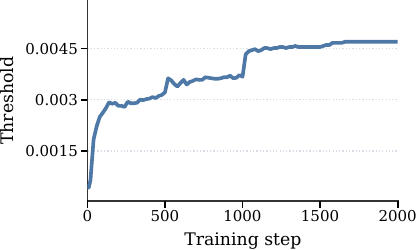}
        \caption{Qwen3-14B on JailbreakBench.}
    \end{subfigure}
    \hfill
    \begin{subfigure}[t]{0.32\textwidth}
        \centering
        \includegraphics[width=\linewidth]{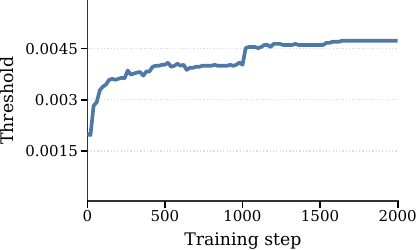}
        \caption{Qwen3-14B on HarmfulBench.}
    \end{subfigure}
    \hfill
    \begin{subfigure}[t]{0.32\textwidth}
        \centering
        \includegraphics[width=\linewidth]{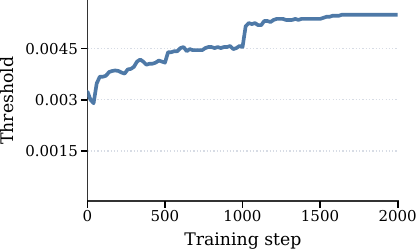}
        \caption{Qwen3-14B on BeaverTails-H.}
    \end{subfigure}

    \vskip\baselineskip

    \begin{subfigure}[t]{0.32\textwidth}
        \centering
        \includegraphics[width=\linewidth]{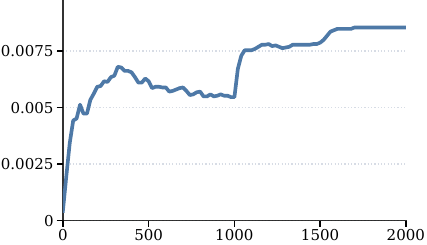}
        \caption{Llama-3.1-8B on JailbreakBench.}
    \end{subfigure}
    \hfill
    \begin{subfigure}[t]{0.32\textwidth}
        \centering
        \includegraphics[width=\linewidth]{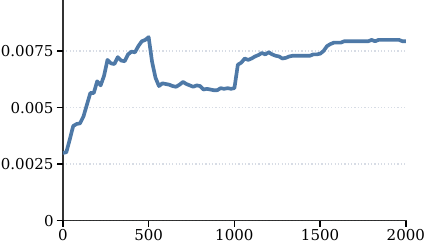}
        \caption{Llama-3.1-8B on HarmfulBench.}
    \end{subfigure}
    \hfill
    \begin{subfigure}[t]{0.32\textwidth}
        \centering
        \includegraphics[width=\linewidth]{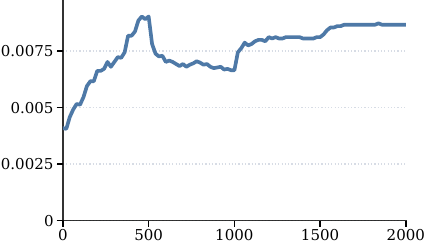}
        \caption{Llama-3.1-8B on BeaverTails-H.}
    \end{subfigure}
    \caption{Threshold trajectories during continued safe training. For each model--dataset pair, we tune the threshold on \(\|hW\|/\|h\|\) at every monitored checkpoint to target a 95\% harmful block rate. Across all six settings, the threshold undergoes an early adjustment phase and then becomes relatively stable later in training.}
    \label{fig:safe_training_threshold}
    \vspace{-8pt}
\end{figure*}

\begin{figure*}[t]
    \centering
    \begin{subfigure}[t]{0.32\textwidth}
        \centering
        \includegraphics[width=\linewidth]{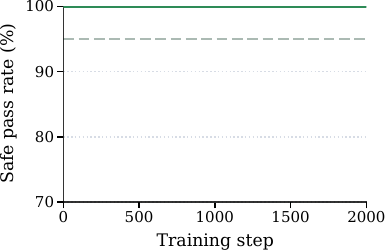}
        \caption{Qwen3-14B on JailbreakBench.}
    \end{subfigure}
    \hfill
    \begin{subfigure}[t]{0.32\textwidth}
        \centering
        \includegraphics[width=\linewidth]{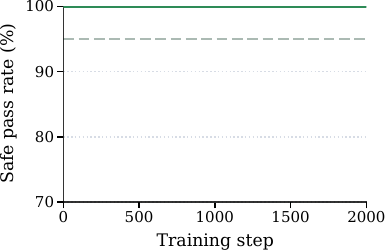}
        \caption{Qwen3-14B on HarmfulBench.}
    \end{subfigure}
    \hfill
    \begin{subfigure}[t]{0.32\textwidth}
        \centering
        \includegraphics[width=\linewidth]{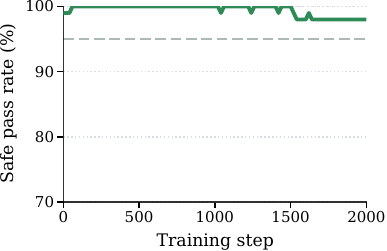}
        \caption{Qwen3-14B on BeaverTails-H.}
    \end{subfigure}

    \vskip\baselineskip

    \begin{subfigure}[t]{0.32\textwidth}
        \centering
        \includegraphics[width=\linewidth]{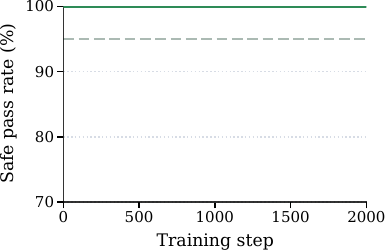}
        \caption{Llama-3.1-8B on JailbreakBench.}
    \end{subfigure}
    \hfill
    \begin{subfigure}[t]{0.32\textwidth}
        \centering
        \includegraphics[width=\linewidth]{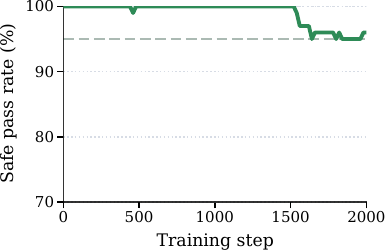}
        \caption{Llama-3.1-8B on HarmfulBench.}
    \end{subfigure}
    \hfill
    \begin{subfigure}[t]{0.32\textwidth}
        \centering
        \includegraphics[width=\linewidth]{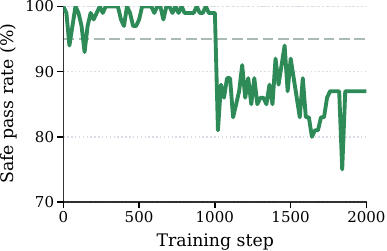}
        \caption{Llama-3.1-8B on BeaverTails-H.}
    \end{subfigure}
    \caption{Safe pass rate during continued safe training under thresholds tuned to maintain a 95\% harmful block rate. The safe pass rate remains at 100\% for all monitored checkpoints in three settings, remains nearly perfect in two additional settings, and stays above 75\% even in the hardest case, showing that safe and harmful samples remain distinguishable under a stringent blocking constraint.}
    \label{fig:safe_training_safe_pass}
    \vspace{-8pt}
\end{figure*}

Figure~\ref{fig:safe_training_threshold} shows that the tuned threshold changes noticeably early in training but becomes relatively stable later across all six model--dataset settings. This suggests that benign adaptation shifts the hidden-state geometry without destroying the operating threshold induced by the cubic layer. Figure~\ref{fig:safe_training_safe_pass} further shows that the corresponding backward gate remains practically useful throughout training: the safe pass rate stays at 100\% for all monitored checkpoints in Qwen3-14B on JailbreakBench and HarmfulBench and in Llama-3.1-8B on JailbreakBench; it remains between 98\% and 100\% for Qwen3-14B on BeaverTails-H and between 95\% and 100\% for Llama-3.1-8B on HarmfulBench; and even in the hardest setting, Llama-3.1-8B on BeaverTails-H, it remains above 75\%. Overall, these results show that safe training unrelated to harmful content does not collapse the alignment into indiscriminate rejection and instead preserves a clear safe--harmful separation under a stringent operating point.

\begin{figure*}[t]
      \centering
      \captionsetup[subfigure]{justification=centering}

      \begin{subfigure}[t]{0.82\textwidth}
          \centering
          \includegraphics[width=0.495\linewidth]{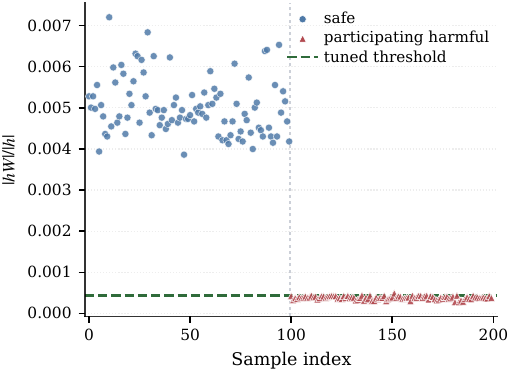}
          \hfill
          \includegraphics[width=0.495\linewidth]{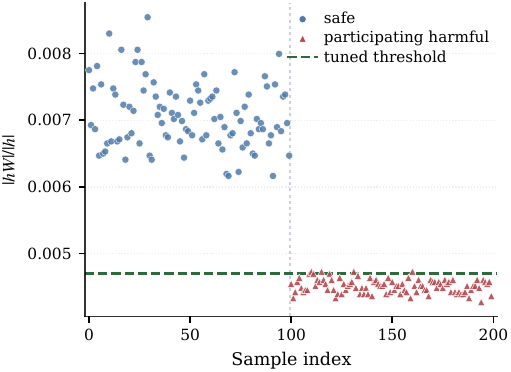}
          \caption{Qwen3-14B on JailbreakBench.}
      \end{subfigure}

      \vspace{0.5em}

      \begin{subfigure}[t]{0.82\textwidth}
          \centering
          \includegraphics[width=0.495\linewidth]{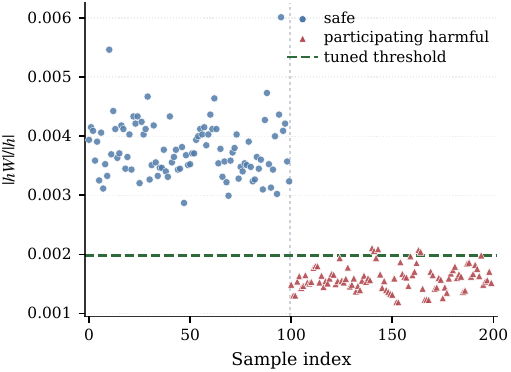}
          \hfill
          \includegraphics[width=0.495\linewidth]{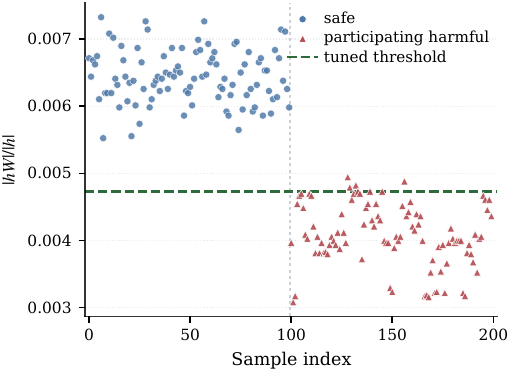}
          \caption{Qwen3-14B on HarmfulBench.}
      \end{subfigure}

      \vspace{0.5em}

      \begin{subfigure}[t]{0.82\textwidth}
          \centering
          \includegraphics[width=0.495\linewidth]{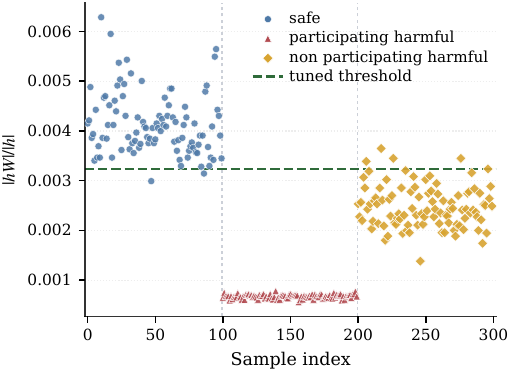}
          \hfill
          \includegraphics[width=0.495\linewidth]{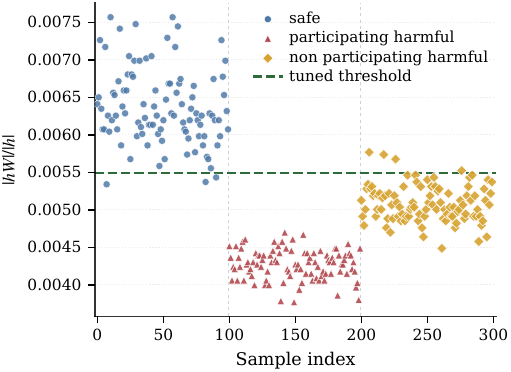}
          \caption{Qwen3-14B on BeaverTails-H.}
      \end{subfigure}

      \caption{Ratio-scatter comparisons for Qwen3-14B before and after continued safe training. Within each row, the left panel shows the model before training and the right panel shows the same model after training completes. Across all three harmful-data settings, the tuned threshold shifts upward during training, but the qualitative separation between safe and harmful samples is largely preserved.}
      \label{fig:safe_training_step_compare_qwen}
      \vspace{-6pt}
  \end{figure*}

\begin{figure*}[t]
      \centering
      \captionsetup[subfigure]{justification=centering}

      \begin{subfigure}[t]{0.82\textwidth}
          \centering
          \includegraphics[width=0.495\linewidth]{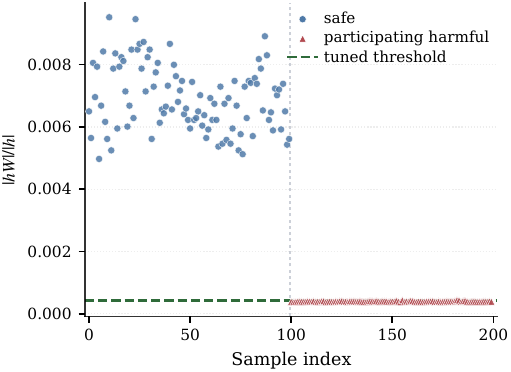}
          \hfill
          \includegraphics[width=0.495\linewidth]{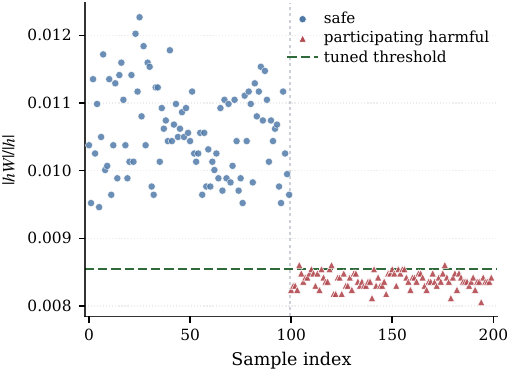}
          \caption{Llama-3.1-8B on JailbreakBench.}
      \end{subfigure}

      \vspace{0.5em}

      \begin{subfigure}[t]{0.82\textwidth}
          \centering
          \includegraphics[width=0.495\linewidth]{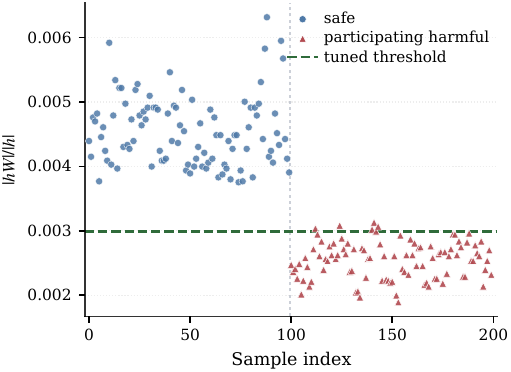}
          \hfill
          \includegraphics[width=0.495\linewidth]{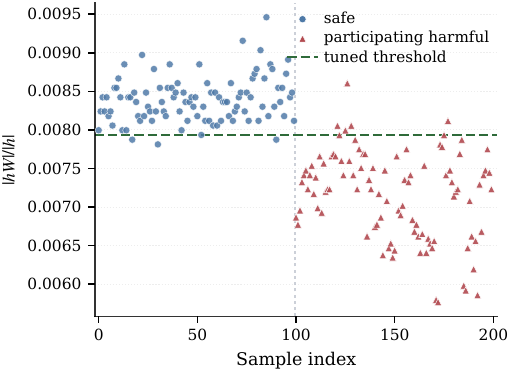}
          \caption{Llama-3.1-8B on HarmfulBench.}
      \end{subfigure}

      \vspace{0.5em}

      \begin{subfigure}[t]{0.82\textwidth}
          \centering
          \includegraphics[width=0.495\linewidth]{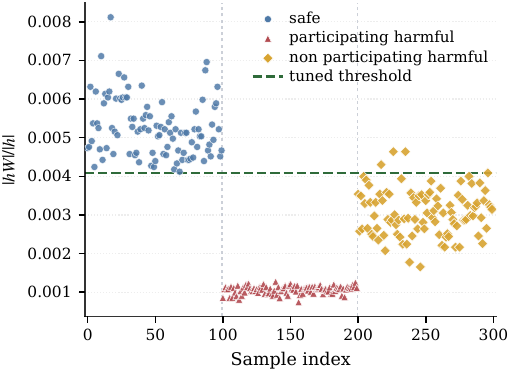}
          \hfill
          \includegraphics[width=0.495\linewidth]{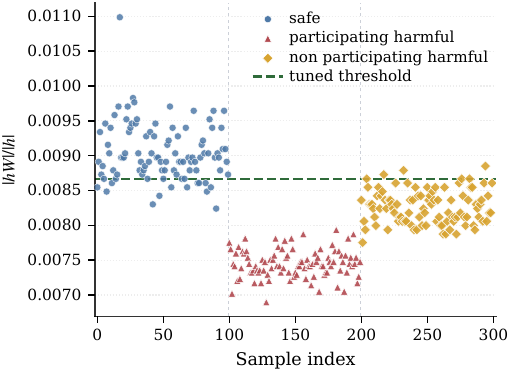}
          \caption{Llama-3.1-8B on BeaverTails-H.}
      \end{subfigure}

      \caption{Ratio-scatter comparisons for Llama-3.1-8B before and after continued safe training. Within each row, the left panel shows the model before training and the right panel shows the same model after training completes. The empirical boundary shifts during training, but the safe--harmful separation remains visible in all three settings, with the margin narrowing most noticeably on BeaverTails-H.}
      \label{fig:safe_training_step_compare_llama}
      \vspace{-6pt}
  \end{figure*}

The ratio-scatter comparisons in Figures~\ref{fig:safe_training_step_compare_qwen} and~\ref{fig:safe_training_step_compare_llama} are consistent with this interpretation. In each row, the left panel shows the model before continued safe training and the right panel shows the same model after training completes under the above configuration. Although the empirical threshold boundary shifts during training, safe samples remain largely above the tuned threshold, whereas harmful samples remain near or below it. Even in the more difficult settings, where the margin narrows, the qualitative separation is preserved. Taken together, these results suggest that continued benign fine-tuning does not by itself destroy the null-space alignment and instead provides a natural calibration phase for fixing a stable release-time threshold.

\section{Cubic Layer Gradient Calculation}
\label{sec:appendix}
\setcounter{equation}{0}
\renewcommand{\theequation}{A\arabic{equation}}
The forward pass of a cubic layer is defined as follows:
\begin{equation}
    \begin{aligned}
        f(h) &= ((hW)h^\mathsf{T})h, \\
&h \in \mathbb{R}^{B \times M \times N},\quad W \in \mathbb{R}^{N \times N}.
    \end{aligned}
\end{equation}

For a given loss function \(L(h)\), we aim to compute \(\frac{\partial L(h)}{\partial h}\). Let 
\begin{equation}
\frac{\partial L(h)}{\partial f(h)} = G_B.
\end{equation}

The differential of \(L\) can be expressed as
\begin{equation}
dL = \operatorname{tr}\left( \left( \frac{\partial L(h)}{\partial h} \right)^\mathsf{T} dh \right) = \operatorname{tr}\left( G_B^\mathsf{T} df \right),
\end{equation}
where \(df = \frac{\partial f}{\partial h} : dh\), with the colon denoting the double dot product (Frobenius inner product).

Set
\begin{equation}
A = hW, \quad B = h^\mathsf{T}, \quad C = h,
\end{equation}
so that
\begin{equation}
f(h) = (AB)C.
\end{equation}

Taking the differential,
\begin{equation}
    \begin{aligned}
        df &= d((AB)C) \\&= d(AB)C + AB\,dC \\&= dA\,BC + A\,dB\,C + AB\,dC.
    \end{aligned}
\end{equation}

Substituting the expressions for \(dA\), \(dB\), and \(dC\):
\begin{equation}
    dA = dhW, \quad dB = dh^\mathsf{T}, \quad dC = dh,
\end{equation}

we obtain
\begin{equation}
    df = (dh W)BC + A\,dh^\mathsf{T} C + AB\,dh.
\end{equation}

Since \(dL = \operatorname{tr}(G_B^\mathsf{T} df)\), it follows that
\begin{equation}
    \begin{aligned}
    dL &= \operatorname{tr}\left( G_B^\mathsf{T} (dh W)BC \right) \\&+ \operatorname{tr}\left( G_B^\mathsf{T} A\,dh^\mathsf{T} C \right) \\&+ \operatorname{tr}\left( G_B^\mathsf{T} AB\,dh \right).
    \end{aligned}
\end{equation}

We now rewrite each term in the form \(\operatorname{tr}\big( (\frac{\partial L}{\partial h})^\mathsf{T} dh \big)\) to extract the gradient.

\textbf{First term}:
\begin{equation}
    \operatorname{tr}\left( G_B^\mathsf{T} (dh W)BC \right) = \operatorname{tr}\left( G_B^\mathsf{T} dh WBC \right).
\end{equation}

Using the cyclic property of the trace, \(\operatorname{tr}(UV) = \operatorname{tr}(VU)\),
\begin{equation}
    \begin{aligned}
    \operatorname{tr}\left( (G_B^\mathsf{T} dh)(WBC) \right) &= \operatorname{tr}\left( (WBC)(G_B^\mathsf{T} dh) \right) \\&= \operatorname{tr}\left( (WBC G_B^\mathsf{T}) dh \right).
    \end{aligned}
\end{equation}

Thus, the contribution to the gradient from the first term is
\begin{equation}
M_1 = (WBC G_B^\mathsf{T})^\mathsf{T} = G_B h^\mathsf{T} h W^\mathsf{T}.
\end{equation}

\textbf{Second term}:
\begin{equation}
\operatorname{tr}\left( G_B^\mathsf{T} A\,dh^\mathsf{T} C \right).
\end{equation}
First, move \(dh^\mathsf{T}\) to the left using cyclic permutations:
\begin{equation}
\operatorname{tr}\left( (G_B^\mathsf{T} A)(dh^\mathsf{T} C) \right) = \operatorname{tr}\left( (dh^\mathsf{T} C)(G_B^\mathsf{T} A) \right).
\end{equation}
Then apply the identity \(\operatorname{tr}(dh^\mathsf{T} M) = \operatorname{tr}(dh M^\mathsf{T}) = \operatorname{tr}(M^\mathsf{T} dh)\) with \(M = C G_B^\mathsf{T} A\). This yields
\begin{equation}
    \begin{aligned}
    \operatorname{tr}\left( dh^\mathsf{T} C G_B^\mathsf{T} A \right) &= \operatorname{tr}\left( dh (C G_B^\mathsf{T} A)^\mathsf{T} \right)
    \end{aligned}
\end{equation}

Therefore,
\begin{equation}
M_2 = C G_B^\mathsf{T} A = h G_B^\mathsf{T} h W.
\end{equation}

\textbf{Third term}:
\begin{equation}
\operatorname{tr}\left( G_B^\mathsf{T} AB\,dh \right)
\end{equation}
is already in the desired form, so
\begin{equation}
\begin{aligned}
 M_3 &= (G_B^\mathsf{T} AB)^\mathsf{T} \\&=  h W^\mathsf{T} h^\mathsf{T} G_B.
\end{aligned}
\end{equation}

Finally, combining all contributions,
\begin{equation}
\frac{\partial L(h)}{\partial h} = M_1 + M_2 + M_3,
\end{equation}
where
\begin{equation}
    \begin{aligned}
    M_1 &= (G_B h^\mathsf{T})(h W^\mathsf{T}), \\\quad M_2 &= (h G_B^\mathsf{T})(h W),\\ \quad M_3 &= h W^\mathsf{T} h^\mathsf{T} G_B.
    \end{aligned}
\end{equation}

\begin{figure*}
    \centering
    \includegraphics[width=0.5\linewidth]{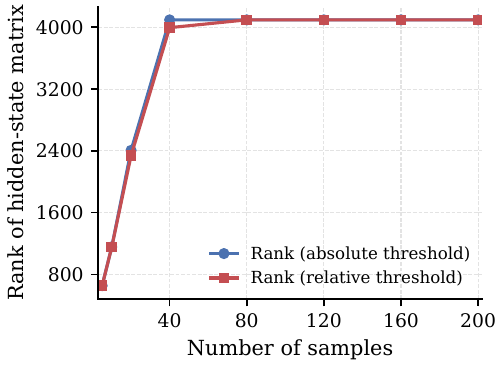}
    \caption{Hidden states matrix rank increases with sample size}
    \label{fig:matrixrank}
\end{figure*}

\section{Details of Inverse Adapter Implementation}\label{appendix:inverse_adapter}

The Inverse Adapter cannot be inserted before the Null Space Cubic Layer, because doing so would alter the layer-input hidden states used by the cubic-layer null-space construction. Therefore, the Inverse Adapter is placed strictly after the Null Space Cubic Layer and serves only as a forward-restoration component.

Its role is distinct from that of the safeguard itself. The Null Space Cubic Layer is the component that realizes gradient blocking for harmful samples inside the protected region. The Inverse Adapter does not contribute additional blocking; instead, it analytically compensates for the deterministic rescaling introduced by the cubic layer so that the released model's outputs remain aligned with those of the base model.

More concretely, for each token position the cubic layer outputs $y=sh$, where $s=hWh^{\mathsf{T}}\ge 0$. Whenever $s>0$, the inverse map is uniquely defined and recovers the original hidden state through
\begin{equation}
h = y (yWy^{\mathsf{T}})^{-1/3}.
\end{equation}
Therefore, whenever the inverse branch is defined, both safe samples and harmful samples can in principle preserve the same forward outputs as the base model, which means preserving both the base model's utility and its original alignment behavior.

In the inverse-adapter-plus-extended-null-space regime studied in this paper, strict zeros are practically absent: exact common-null-space directions are replaced by smallest-singular-value directions, and finite-precision computation leaves suppressed but non-zero activations rather than mathematically exact zeros for almost all samples. As a result, the analytic inverse remains applicable to essentially all samples in our evaluated setting, so the Inverse Adapter still receives valid inputs outside the thresholded blocked regime and the forward behavior of both safe and harmful inputs is preserved up to negligible numerical error. Under this interpretation, the practical utility question is dominated by whether benign data remain outside the protected region rather than by inverse reconstruction error itself. The cubic layer is responsible for deciding which samples have their gradients blocked, while the Inverse Adapter is responsible for keeping the forward pass close to the original base-model computation.

\section{A Preliminary Parameter-Fusion Instantiation of the Protected-Release Assumption}
\label{appendix:parameter_fusion}

The main paper studies the blocking mechanism itself under the PPOW setting. As one feasible hardening path for this assumption, we further explore parameter fusion, in which the explicit safeguard-facing computation is absorbed into downstream Transformer computation to reduce the extent to which the released artifact exposes a locally detachable module. Importantly, the goal of fusion is not to improve the blocking mechanism itself, but to provide a more entangled release form that may raise the cost of naive structural removal.

Concretely, our preliminary fusion path targets the computation immediately downstream of the cubic layer: we treat the explicit Inverse Adapter together with the following Transformer block as the teacher-side protected computation, and train a single fused Transformer-like block to reproduce their combined behavior. In this way, the released artifact no longer needs to expose the original explicit safeguard-facing interface in modular form. If the fused block has already co-adapted to the safeguarded hidden-state manifold, then simply deleting the explicit safeguarded transformation is no longer a purely local edit, because the downstream fused computation itself has been fit against that safeguarded geometry.

At the same time, this study should be interpreted as a preliminary feasibility check rather than as a fully optimized hardening pipeline. In our current setting, the fused block is fitted with limited calibration/distillation data and limited optimization budget, so it does not yet cover the full hidden-state manifold induced by the safeguarded model. As a result, the current fusion quality should not be read as the best achievable fidelity--hardening trade-off; rather, it provides an initial demonstration that the protected computation can in principle be absorbed into downstream Transformer computation while still motivating the protected-release threat model used in the main paper.

\begin{table*}[t]
\centering
\begin{tabular}{l c c c c}
\hline
\textbf{Model} & \textbf{Dataset Variant} & \textbf{Threshold Selection} & {Safe Pass Rate} &{Harmful Block Rate} \\
\hline
Qwen & JailbreakBench & 0.00049 &100.0\% & 100.0\%\\
Qwen & HarmfulBench & 0.0021 &100.0\% & 100.0\% \\
Qwen & BeaverTails & 0.0036 &79.0\% &100.0\% \\
\hline
Llama & JailbreakBench & 0.00044 &100.0\% &100.0\% \\
Llama & HarmfulBench & 0.0031 &100.0\% &100.0\% \\
Llama & BeaverTails & 0.0046 & 69.0\% & 100.0\%\\
\hline
\end{tabular}
\caption{Calibration Thresholds in the 100\%-Block Experimental Setting}
\label{threshold_selection}
\end{table*}

\section{Extended Null Space Selection}\label{sec:appendix_impl}
As shown in Figure \ref{fig:matrixrank}, when the number of harmful samples used to build $S_h$ increases, the stacked hidden-state matrix tends to become full rank in practice, so $\operatorname{null}_{\text{common}}(S_h)$ can become trivial and the strict constraint $hW=hW^\mathsf{T}=0$ may have no non-zero solution. To keep the same objective while avoiding this rank-collapse issue, we replace strict common-null-space selection with a smallest-singular-direction approximation. 
Concretely, let $X$ denote the stacked matrix formed from vectors in $S_h$, and compute $X=U\Sigma V^{\mathsf{T}}$. Instead of taking basis vectors from the exact null space (singular value exactly zero), we take the right singular vectors associated with the smallest singular values to form $V_{\text{small}}$, and set $W=V_{\text{small}}V_{\text{small}}^{\mathsf{T}}$ (or its symmetric equivalent) so that, for $h\in S_h$, $\|hW\|$ and $\|hW^{\mathsf{T}}\|$ are minimized rather than forced to be exactly zero. This yields an extended null-space criterion that is numerically stable under large-sample, near-full-rank conditions, while preserving the gradient-blocking mechanism through small residual factors in the cubic-layer gradient terms.

\subsection{Sample Limit for Exact Null-space Selection}

For exact null-space selection, the key question is how many harmful training examples can be incorporated before the shared null space collapses. 
To measure this limit, we construct harmful hidden-state matrices from the exact same prompt/response formatting used in the null-space pipeline for each model--dataset pair. 
For every harmful example, we run the model forward, collect the layer-input hidden states for the target Cubic layer, and stack the token-level hidden states of the first $n$ harmful samples into a single matrix.
We then perform exact null-space extraction on this stacked matrix and increase $n$ until the extracted null space becomes empty. 
We define the \emph{sample limit} as the largest number of harmful samples for which the exact construction still yields a non-trivial null space, and the corresponding \emph{token limit} as the cumulative number of tokens contained in those samples. In this way, the table reports the maximal harmful-data scale under which exact null-space selection remains feasible.

Geometrically, this behavior is expected. As more harmful samples are added, the stacked hidden-state matrix spans a progressively larger subspace and its rank correspondingly increases. Once the harmful hidden states become sufficiently diverse, the intersection of their orthogonal complements rapidly shrinks, so the common null space approaches zero dimension. At that point, the exact symmetric matrix construction satisfying $hW=hW^{\mathsf{T}}=0$ becomes trivial or disappears entirely. Therefore, the sample limit for exact null-space selection is fundamentally governed by rank growth and null-space degeneracy: beyond a certain harmful-sample scale, the exact construction can no longer provide a meaningful non-zero solution.

Table~\ref{tab:exact_nullspace_limit} shows that this feasible region is in fact quite small for both model families. First, the exact construction supports only tens of harmful samples rather than hundreds, confirming that exact null-space selection is intrinsically capacity-limited. Second, the limit depends strongly on dataset diversity: HarmfulBench reaches the limit the fastest for both Qwen and Llama, indicating that its harmful hidden states span the relevant subspace more efficiently and thus eliminate the shared null space with fewer examples. By contrast, JailbreakBench permits the largest exact null-space construction, while BeaverTails lies in between. Third, the same trend is reflected in token space: even when measured by cumulative token count instead of sample count, the feasible region remains narrow, typically only a few thousand to roughly ten thousand tokens. Overall, these results show that exact null-space selection is structurally fragile at scale, which motivates moving to extended / approximate null-space constructions when broader harmful coverage is required.

\begin{table*}[t]
  \centering
  \begin{tabular}{l c c c}
  \hline
  \textbf{Model} & \textbf{Dataset}  &\textbf{Sample Limit} & \textbf{Token Limit}\\
  \hline
  Qwen  & JailbreakBench    & 91              & 11744 \\
  Qwen  & HarmfulBench      & 14              & 5503 \\
  Qwen  & BeaverTails       & 45              & 8067 \\
  \hline
  Llama & JailbreakBench    & 73              & 9239 \\
  Llama & HarmfulBench      & 13              & 4671 \\
  Llama & BeaverTails       & 35              & 6256 \\
  \hline
  \end{tabular}
  \caption{Maximum harmful-data scale at which exact null-space selection remains non-trivial. For each model--dataset pair, we increase the number of harmful samples used to build the stacked hidden-state matrix and report the largest sample count for which the exact null space is still non-empty. Token limit reports the corresponding cumulative token count of those harmful samples. The small limits across all settings show that exact null-space selection quickly degenerates as harmful hidden-state diversity increases.}
  \vspace{-10pt}
  \label{tab:exact_nullspace_limit}
\end{table*}

\subsection{Sample Limit for Extended Null Space Selection}
For extended null-space selection, the exact null-space degeneracy described above is alleviated by replacing zero-singular-value directions with smallest-singular-value directions. As a result, its sample limit is no longer best defined by whether the exact symmetric matrix construction exists, but rather by how much useful safe behavior can still be preserved under a fixed safety requirement.

Accordingly, we adopt a utility-oriented operational definition of sample limit for the extended setting. Specifically, we first determine a build-time threshold that enforces a 95\% harmful block rate on the participating harmful samples and then evaluate the resulting held-out safe pass rate. This metric asks a concrete question: as more harmful samples are used to construct the extended null space, how much safe-data utility can still be retained while maintaining a stringent safety target?

Under this definition, larger harmful build sets are not automatically better. If the threshold chosen to satisfy the build-time 95\% harmful block criterion leads to a declining held-out safe pass rate, then the usable capacity of the construction is approaching its limit. We therefore use the safe-pass-rate curve under the fixed 95\% safety target to characterize the sample limit of extended null-space selection, and compare it against exact null-space selection to determine which method better delays the onset of performance degradation. We additionally report the corresponding held-out harmful block rate to verify that the lower ratio tendency also extends beyond the build set strongly enough to support threshold-based blocking on held-out harmful data.

\begin{figure*}[!htbp]
    \centering
    \includegraphics[width=0.5\linewidth]{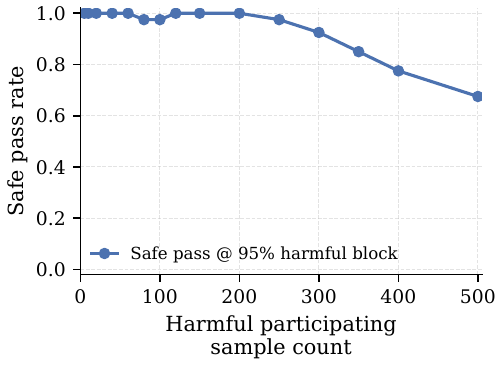}
    \caption{Safe sample pass rate (with 95\% harmful block rate) as a function of the number of participating harmful samples on the BeaverTails dataset, evaluated with the Qwen3-14B model. The extended null space method maintains 100\% safe pass rate at 200 samples and over 60\% at 500 samples, substantially outperforming Exact Null-space Selection.}
    \label{fig:safe_pass_rate}
\end{figure*}

Given the limited sample sizes of JailbreakBench and HarmfulBench, we demonstrate the sample capacity of the extended null space method using the Qwen3-14B model on the BeaverTails dataset. The threshold for each method is selected to achieve a 95\% harmful block rate while maximizing the safe sample pass rate. As shown in Figure~\ref{fig:safe_pass_rate}, the extended null space method maintains a 100\% safe pass rate with approximately 200 samples, and still preserves over 60\% safe pass rate even with 500 samples. In this evaluated BeaverTails/Qwen3-14B setting, this indicates markedly higher sample capacity than the Exact Null-space Selection approach.

\section{Frozen-Threshold Stability under Incomplete OOD Coverage}
\label{appendix:ood_coverage_stability}

The main paper focuses on a strict frozen-threshold release regime in which the calibrated threshold fully blocks the relevant harmful reference set used for deployment-style evaluation. Here we study a more conservative boundary case on Qwen3-14B with BeaverTails-H: the frozen threshold is fixed from the strict 100\% coverage reference run, but the subsequent continued training data are benign yet distribution-shifted with respect to the release-time calibration set. This setting lets us test whether the same threshold-controlled gate remains stable under such adaptation, without any online threshold retuning.

Table~\ref{tab:ood_coverage_stability_summary} summarizes the final monitored operating points. The strict 100\% reference run remains stable throughout training, with safe pass rate 0.79 and harmful block rate 1.00 at the final checkpoint. Under incomplete OOD coverage with learning rate $1\times10^{-5}$, the behavior remains nearly unchanged, ending at safe pass rate 0.78 and harmful block rate 1.00. At learning rate $4\times10^{-5}$, the harmful block rate still remains at 1.00, but the safe pass rate decreases to 0.59, indicating that the backward gate induced by the frozen threshold remains active but becomes more conservative on safe samples. By contrast, at learning rate $3\times10^{-4}$, the final safe pass rate rises to 1.00 while the harmful block rate drops to 0.00, showing that the original harmful--safe separation no longer remains stable in this high-update regime.

\begin{figure*}[t]
    \centering
    \begin{subfigure}[t]{0.48\textwidth}
        \centering
        \includegraphics[width=\linewidth]{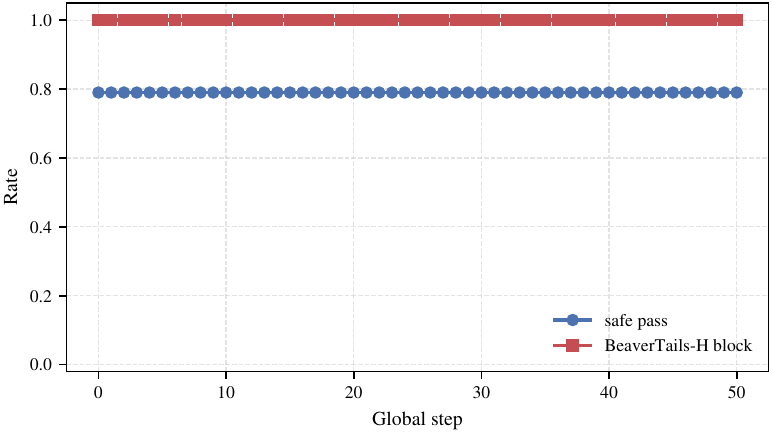}
        \caption{Strict100 reference.}
    \end{subfigure}
    \hfill
    \begin{subfigure}[t]{0.48\textwidth}
        \centering
        \includegraphics[width=\linewidth]{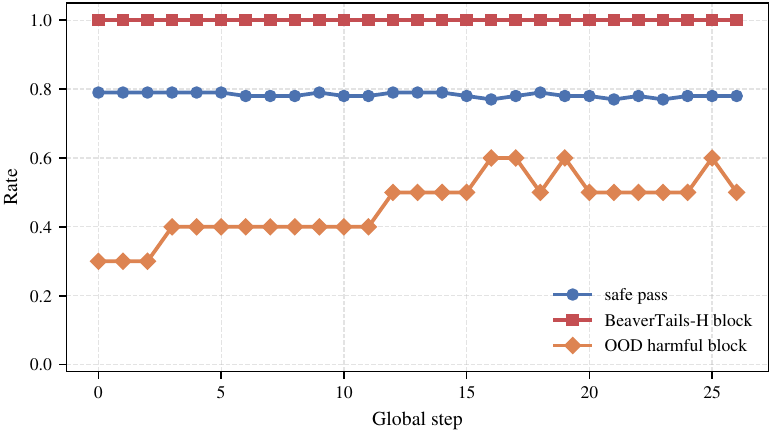}
        \caption{Frozen threshold, $\mathrm{lr}=1\times10^{-5}$.}
    \end{subfigure}

    \vskip\baselineskip

    \begin{subfigure}[t]{0.48\textwidth}
        \centering
        \includegraphics[width=\linewidth]{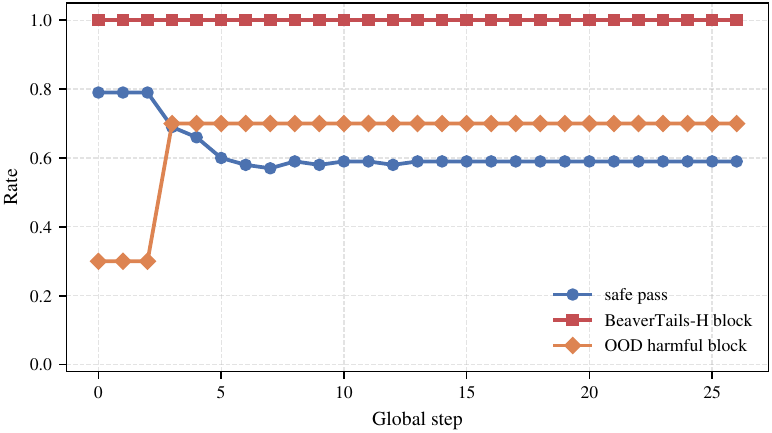}
        \caption{Frozen threshold, $\mathrm{lr}=4\times10^{-5}$.}
    \end{subfigure}
    \hfill
    \begin{subfigure}[t]{0.48\textwidth}
        \centering
        \includegraphics[width=\linewidth]{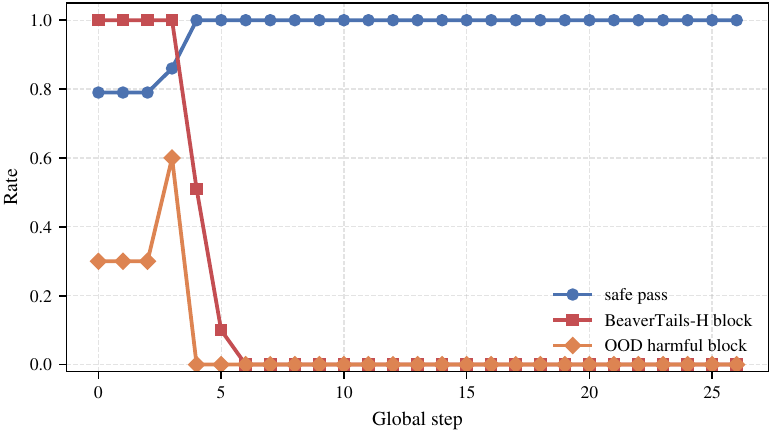}
        \caption{Frozen threshold, $\mathrm{lr}=3\times10^{-4}$.}
    \end{subfigure}
    \caption{Safe-pass and harmful-block trajectories under a fixed strict-100 release threshold. The strict reference run and the $1\times10^{-5}$ OOD run remain close to their initial operating points, the $4\times10^{-5}$ run keeps harmful blocking but sacrifices safe pass rate, and the $3\times10^{-4}$ run rapidly loses harmful blocking.}
    \label{fig:ood_coverage_stability_curves}
    \vspace{-6pt}
\end{figure*}

\begin{figure*}[t]
    \centering
    \captionsetup[subfigure]{justification=centering}

    \begin{subfigure}[t]{0.82\textwidth}
        \centering
        \includegraphics[width=\linewidth]{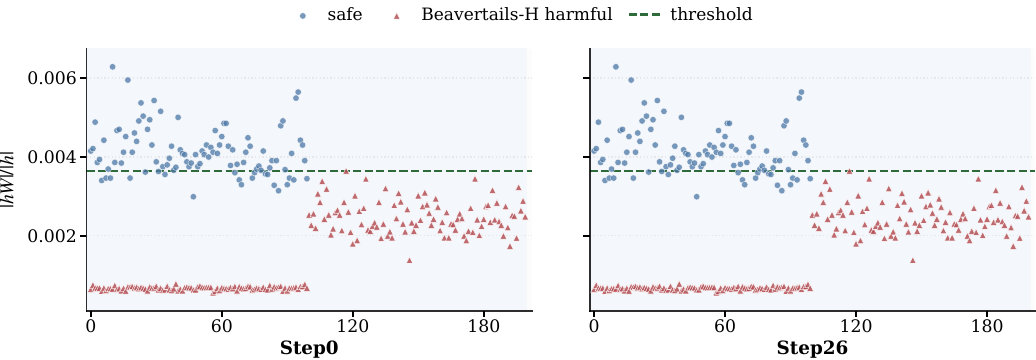}
        \caption{Strict100 reference.}
    \end{subfigure}

    \vspace{0.5em}

    \begin{subfigure}[t]{0.82\textwidth}
        \centering
        \includegraphics[width=\linewidth]{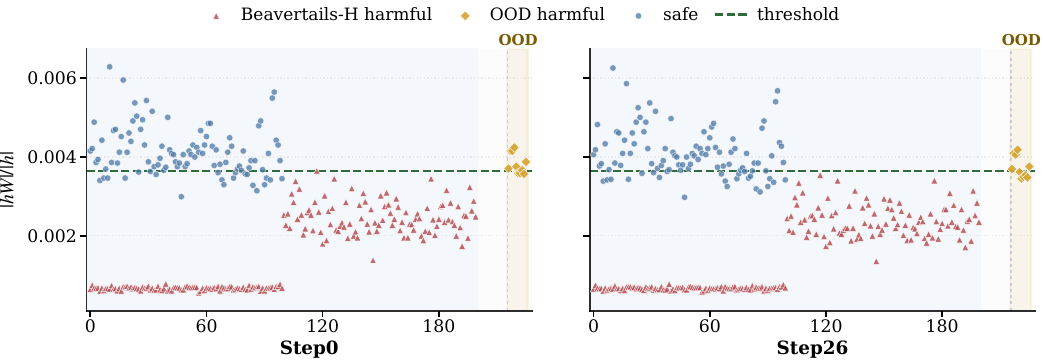}
        \caption{Frozen threshold, $\mathrm{lr}=1\times10^{-5}$.}
    \end{subfigure}

    \vspace{0.5em}

    \begin{subfigure}[t]{0.82\textwidth}
        \centering
        \includegraphics[width=\linewidth]{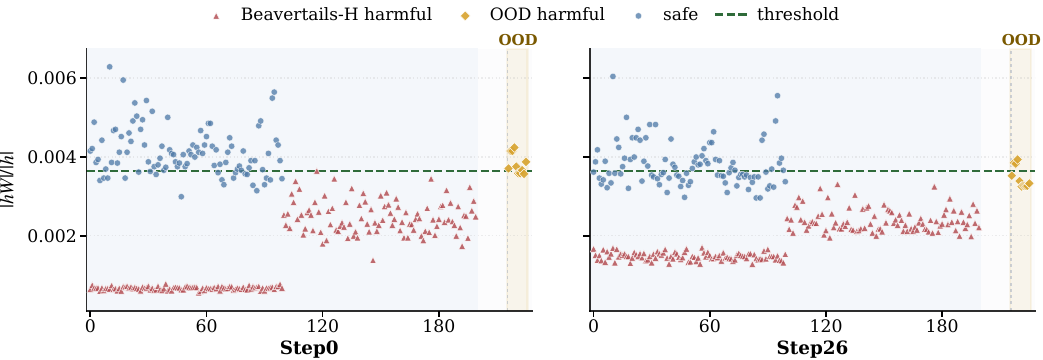}
        \caption{Frozen threshold, $\mathrm{lr}=4\times10^{-5}$.}
    \end{subfigure}

    \vspace{0.5em}

    \begin{subfigure}[t]{0.82\textwidth}
        \centering
        \includegraphics[width=\linewidth]{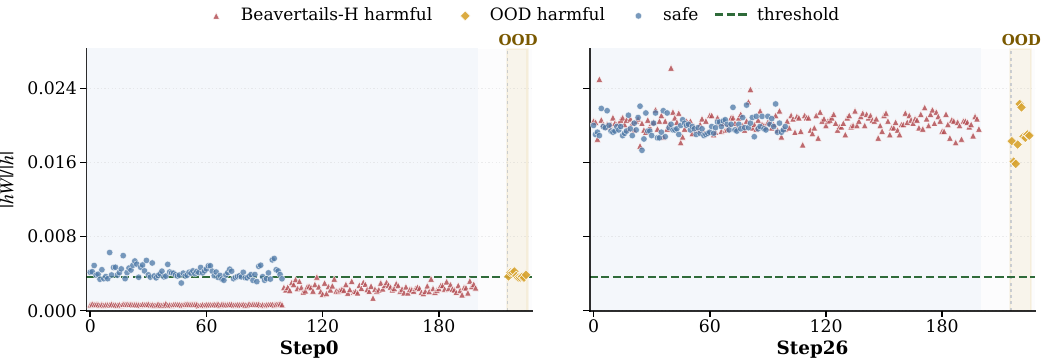}
        \caption{Frozen threshold, $\mathrm{lr}=3\times10^{-4}$.}
    \end{subfigure}

    \caption{Ratio-scatter comparisons at initialization and the final monitored step under a fixed strict-100 release threshold. The strict reference and $1\times10^{-5}$ runs preserve the original safe--harmful separation, the $4\times10^{-5}$ run pushes safe samples toward the threshold boundary, and the $3\times10^{-4}$ run collapses the separation so that the frozen threshold no longer cleanly separates harmful from safe samples.}
    \label{fig:ood_coverage_ratio_pairs}
    \vspace{-6pt}
\end{figure*}

Figure~\ref{fig:ood_coverage_stability_curves} shows the corresponding dynamics over training. The strict 100\% reference run is flat across all monitored checkpoints, and the $1\times10^{-5}$ run stays close to that reference trajectory. The $4\times10^{-5}$ run preserves harmful blocking but gradually shifts toward lower safe pass rate, whereas the $3\times10^{-4}$ run rapidly loses harmful blocking after only a few monitored steps. Figure~\ref{fig:ood_coverage_ratio_pairs} is consistent with this interpretation: the ratio-scatter geometry remains visually similar for the strict reference and $1\times10^{-5}$ runs, compresses toward the threshold boundary at $4\times10^{-5}$, and collapses at $3\times10^{-4}$ so that the frozen threshold no longer cleanly separates harmful from safe samples.

Taken together, these results support a conservative interpretation of fixed-threshold deployment. When the release-time threshold is calibrated with complete harmful coverage, the corresponding backward gate can remain stable under mild continued training. When continued adaptation is OOD with respect to the release-time calibration set, however, stability becomes update-dependent: small updates can preserve the threshold-controlled gate, intermediate updates can make it more conservative, and sufficiently large updates can erase the original harmful--safe separation.

\begin{table*}[t]
\centering
\small
\setlength{\tabcolsep}{6pt}
\begin{tabular}{lccc}
\toprule
\textbf{Setting} & \textbf{LR} & \textbf{Final Safe Pass} & \textbf{Final Harmful Block} \\
\midrule
threshold policy & -- & 0.79 & 1.00 \\
Frozen threshold & $1\times10^{-5}$ & 0.78 & 1.00 \\
Frozen threshold & $4\times10^{-5}$ & 0.59 & 1.00 \\
Frozen threshold & $3\times10^{-4}$ & 1.00 & 0.00 \\
\bottomrule
\end{tabular}
\caption{Final monitored operating points on Qwen3-14B with BeaverTails-H under a fixed strict-100 release threshold. Mild benign distribution shift preserves the reference operating point, moderate learning rate reduces safe pass while keeping full harmful blocking, and aggressive learning rate destroys the original harmful--safe separation.}
\label{tab:ood_coverage_stability_summary}
\end{table*}

\end{document}